\documentclass[10pt,aps,prd,a4paper,nofootinbib,
preprintnumbers,superscriptaddress,showpacs,showkeys,twocolumn]{revtex4}

\usepackage{amsmath}
\usepackage{amssymb}
\usepackage{bm}
\usepackage{graphicx}
\usepackage{xcolor}
\usepackage{slashed}
\usepackage{comment}
\usepackage[normalem]{ulem}

\newcommand{\be}{\begin{equation}}
\newcommand{\ee}{\end{equation}}

\begin{document}

\title{Quantum decoherence: a study applied to quarkonium-like bound states in strongly interacting matter}

\author{Gabriele Coci} \email{gabriele.coci@dfa.unict.it}
\affiliation{Department of Physics and Astronomy "Ettore Majorana", University of Catania, Via Santa Sofia 64, I-95123 Catania, Italy}\affiliation{INFN-LNS Laboratori Nazionali del Sud, Via S. Sofia 62, I-95123 Catania, Italy}

\author{Salvatore Plumari} \email{salvatore.plumari@dfa.unict.it}
\affiliation{Department of Physics and Astronomy "Ettore Majorana", University of Catania, Via Santa Sofia 64, I-95123 Catania, Italy}\affiliation{INFN-LNS Laboratori Nazionali del Sud, Via S. Sofia 62, I-95123 Catania, Italy}

\author{Giuseppe Falci} \email{giuseppe.falci@dunict.it}
\affiliation{Department of Physics and Astronomy "Ettore Majorana", University of Catania, Via Santa Sofia 64, I-95123 Catania, Italy}\affiliation{INFN Sezione di Catania, Via S. Sofia 64, I-95123 Catania, Italy}

\begin{abstract}
We study the quantum decoherence of a bound state interacting with a reservoir of strongly interacting matter within the framework of open quantum systems. The bound state is modeled as a quantum harmonic oscillator whose parameters are tuned to reproduce the root-mean-square radius of $J/\Psi$ particle.
The surrounding medium, representing the many degrees of freedom of strongly interacting matter, acts as an environment that induces dissipation and decoherence through system–reservoir coupling. By analyzing the time evolution of the reduced density matrix, we quantify the loss of quantum coherence and its dependence on medium properties. 
Subsequently, we extend the model by introducing a time dependence in the system–thermal bath coupling, thereby simulating a temperature evolution similar to that occurring during the expansion of a fireball in the central region of heavy-ion collisions. 
We find that a temperature evolution has a relevant impact on the way the system loses coherence through the coupling with the expanding medium.
Finally, we estimate the impact of the time-dependent temperature on the decoherence process, also analyzing a scenario that includes viscous effects without finding a significant change with respect to ideal hydrodynamical evolution.

\end{abstract}


\maketitle
\section{Introduction}

Ultra-relativistic heavy-ion collisions provide a unique opportunity to study strongly interacting matter under extreme conditions of temperature and energy density, where Quantum Chromodynamics (QCD) predicts the formation of a deconfined phase known as the quark–gluon plasma (QGP). Understanding the dynamical properties of this medium and its microscopic degrees of freedom remains one of the central goals of contemporary high-energy nuclear physics.

Heavy quarkonium states, such as charmonium and bottomonium, play a special role in this context. Owing to the large masses of heavy quarks, these bound states are predominantly produced in the early stages of the collision and subsequently propagate through the evolving medium. Their survival probability is sensitive to color screening, in-medium dissociation, and recombination processes, making quarkonium suppression a powerful probe of deconfinement and QGP properties (see reviews~\cite{Rapp:2009my,Rothkopf:2019ipj,Dong:2019unq,Zhao:2020jqu}).
The original idea that quarkonium suppression could serve as a signature of QGP formation was proposed by Tetsuo Matsui and Helmut Satz~\cite{Matsui:1986dk}. Since then, extensive experimental measurements at the CERN SPS~\cite{NA50:1996lag,NA60:2006ncq}, at RHIC~\cite{PHENIX:2006gsi,STAR:2009irl,STAR:2013kwk}, and LHC~\cite{ALICE:2012jsl,ALICE:2018bdo,CMS:2012gvv} (Ref.~\cite{Andronic:2015wma} for report review) have provided high-precision data on quarkonium production in nucleus–nucleus collisions. These observations reveal a complex interplay between suppression mechanisms, regeneration effects, and cold nuclear matter contributions.
From the theoretical standpoint, several approaches have been developed to describe in-medium quarkonium dynamics, including potential models driven by lQCD inferences~\cite{Laine:2006ns,Petreczky:2008px}, effective field theory frameworks~\cite{Brambilla:1999xf}, transport equations~\cite{Grandchamp:2002wp,Song:2023zma}, and open quantum system methods~\cite{Akamatsu:2011se,Akamatsu:2012vt,Akamatsu:2014qsa,Brambilla:2016wgg,Brambilla:2017zei,Brambilla:2020qwo,Yao:2020xzw,Yao:2025jyx,Blaizot:2015hya,Katz:2015qja} (see Ref.~\cite{Andronic:2024oxz} for a comparative study among several of these approaches).

The theory of open quantum systems (OQS)~\cite{BreuerPetruccione2002} has been shown to provide a valid and powerful theoretical framework for studying a wide range of problems not only in its natural condensed matter framework, but also in high-energy physics. Some examples include thermalization of nuclear bound states~\cite{Rais:2025fps,Neidig:2023kid}, jets~\cite{DeJong:2020riy,Mehtar-Tani:2025xxd}, entropy production~\cite{Coci:2025drb,Iida:2014wea} and quarkonium suppression~\cite{Borghini:2011ms,Akamatsu:2014qsa,Katz:2015qja,Brambilla:2016wgg,Brambilla:2020qwo,Blaizot:2017ypk,Delorme:2024rdo}.

The central idea of the open quantum systems framework is to derive a master equation governing the evolution of the density matrix of a quantum system that interacts with a larger environment, typically referred as a reservoir. The coupling between the system and its surroundings produces characteristic phenomena that stem from the intrinsically quantum nature of the dynamics, most notably decoherence.
In this context, quantum decoherence describes the mechanism through which a system progressively loses the coherence of its superposition states as a consequence of environmental interactions. Unlike an idealized isolated system, an open quantum system continuously exchanges energy and information with external degrees of freedom. Through this interaction, quantum phase correlations become entangled with the environment and effectively inaccessible to local observations. Consequently, interference effects are suppressed and the system exhibits behavior that resembles classical physics. Importantly, decoherence does not imply a physical collapse of the wavefunction; rather, it provides a dynamical explanation for the emergence of classical properties from quantum systems due to environmental coupling~\cite{Schlosshauer:2003zy,Schlosshauer:2019ewh,Zurek:2003zz,Zurek:1991vd}.

The derivation of a quantum master equation for the density matrix of a subsystem interacting with an environment with dissipation and decoherence relies on a set of physical approximations. 
The GKSL master equation also known as Lindblad equation~\cite{Gorini:1975nb,Lindblad:1975ef} represents the most general class of real-time dynamical Markovian evolution equation for the density matrix of an open system that preserves its fundamental properties, namely complete positivity and unit trace. Currently there are two physical regimes where Lindblad equations are derivable: the optical limit and the quantum Brownian motion.
The distinction among the two regimes arises from different hierarchies of characteristic times: the environment correlation time $\tau_E$, the intrinsic system time $\tau_S$ and the relaxation time of the system-environment interaction $\tau_R$. In both regimes, the Markovian approximation requires the decay of environment/bath correlations to occur much faster than the typical time the system suffers a significant change by losing energy, i.e. $\tau_E \ll \tau_R$. Hence, memory effects can be neglected.

In the optical regime the intrinsic time of the system is such that $\tau_S \ll \tau_R$. In terms of energy relation, this is equivalent to saying that the characteristic frequency of the system $\omega_0 \sim \tau_S^{-1}$ is much larger than the damping rate induced by the environment interaction, which usually goes as $T^2/M \sim \tau_R^{-1}$, being $M$ the mass of the system and $T$ the temperature of the reservoir, assumed at thermal equilibrium. In quantum optics such conditions allow to perform a rotating wave approximation, which means that within the interaction Hamiltonian the highly oscillating terms are neglected and that transitions are induced only by environment modes resonating with the system frequency $\omega_0$. In heavy-ion collisions (HICs), heavy quarks
are characterized by a mass $M \gg T$ and $M \gg \Lambda_{QCD}$.
Therefore, heavy quark bound states, known as quarkonia, are non-relativistic and their suppression in strongly interacting matter depends on their binding energy $E_B$ which is modified by complex in-medium effects. In this context, the quantum optical regime is applicable to quarkonia states which are tightly bound $E_B \sim \omega_0 \simeq T$~\cite{Borghini:2011ms}, in particular to the ground state of charmonium ($J/\Psi$) and bottomonium ($Y(1S)$). 

At higher temperatures, when $E_B \ll T$, the suppression of quarkonia in strongly interacting matter has been investigated using the quantum Brownian motion framework~\cite{Brambilla:2016wgg,Brambilla:2020qwo,Brambilla:2022ynh,Miura:2019ssi,Delorme:2024rdo}. Under the assumption that thermal fluctuations of the medium dominate over the intrinsic binding scale of the state, the relation among binding energy and temperature is such that the intrinsic time of the system is $\tau_S \sim E_B^{-1} \gg T^{-1} \sim \tau_E$, the bath correlation time. This corresponds precisely to the standard OQS approximation to derive a quantum master equation for dissipative-stochastic dynamics characteristic of the Brownian regime, i.e. the Caldeira-Leggett model~\cite{Caldeira:1982iu,caldeira1983quantum}.

In this paper we address the problem to study quantum decoherence for quarkonium states by means of the Lindblad equation in the optical regime~\cite{Borghini:2011ms}. In order to do this, we employ a well known model in condensed matter physics which consists of a one-dimensional damped harmonic oscillator (H.O.) coupled to a reservoir of independent bosonic fields in thermal equilibrium.~\cite{carmichael1993quantum}. 
In particular, the system frequency $\omega_0$ is tuned to reproduce the $J/\Psi$ 
root-mean-square radius from Particle Data Group~\cite{ParticleDataGroup:2024cfk}. 
The system-reservoir interaction is modelled as an amplitude-coupling among the quarkonium ladder operators and the bath operators whose strength is phenomenologically inferred from quarkonium in medium-dissociation widths calculated from a non-perturbative microscopic many-body theory~\cite{Grandchamp:2002wp,Rapp:2009my}.

In particular, we focus on the possibility to extend this model by introducing an additional "hydrodynamical" time-scale coming from the QGP expansion. In particular, this time-scale falls in between the the large relaxation time of the system and the fast decay time of reservoir correlations, allowing to introduce a temperature dependent system-reservoir coupling strength. 
Our main goal is to provide within an OQS framework a resonable estimate about the time-scale at which the quarkonium wave-function loses coherence due to the presence of a static and an expanding QGP medium~\cite{Akamatsu:2011se,Kajimoto:2017rel}. A short decoherence time would signal the fast process which drives the system towards a semi-classical description both during its evolution and above all at the hadronization.

This paper is organized as follows: after this Introduction, in Sec.~\ref{sec:II} we describe the quantum master equation for the damped H.O. coupled to a heat bath at fixed temperature $T$ in amplitude-coupling interaction. Then, we discuss about a possible coupling of the system to an evolving environment characterized by a time-dependent temperature. Still in Sec.~\ref{sec:II} we describe the set up of the model parameters as well as the numerical solution of the Lindblad equation. In Sec.~\ref{sec:III} we discuss about the thermalization of ground state population in comparison with semi-classical rate equations, while in Sec.~\ref{sec:IV} we analyze quantum decoherence process induced by the static heat bath. In Sec.~\ref{sec:V} we describe the same phenomenon projecting the quantum master equation in the phase-space representation, then showing the evolution of the Wigner distribution and the "classicalization" of the system driven by dissipation and decoherence. Finally, in Sec.~\ref{sec:VI} we present our results for quantum decoherence of quarkonium coupled to an environment which expands along one-dimension according to ideal hydrodynamics, showing also a case where viscous effects are included. We finish this article with a short summary and the conclusions.
In this work we use natural units $\hbar = k_B = c = 1$. 

\section{Model}
\label{sec:II}
\subsection{Lindblad equation in the optical regime}
In this section we briefly review the derivation of the Lindblad master equation for a damped harmonic oscillator (H.O.) linearly coupled to reservoir which is composed of a continuum of harmonic oscillators~\cite{carmichael1993quantum,BreuerPetruccione2002}. The full Hamiltonian of the model is written as
$ H = H_S + H_R + H_{SR} $
where
\begin{align}
    H_S &= \omega_0 a^\dagger a \\
    H_R &= \sum_j \omega_j b_j^\dagger b_j  \\
    H_{SR} &= \sum \left( \kappa_j b_j a^\dagger + \kappa_j^* b_j^\dagger a \right)
    \label{eq:Hamplitudecoupling}
\end{align}
The system $S$ is a one-dimensional quantum H.O. with frequency $\omega_0$ whose ladder operators are $a$ and $a^\dagger$. 
The reservoir $R$ is a collection of indipendent harmonic oscillators with characteristic frequencies $\omega_j$ and creation and destruction operators $b_j^\dagger$, $b_j$.
The term $H_{SR}$ describes the interaction between the system and each mode of the reservoir whose strength is controlled by the complex coefficients $\kappa_j$.
The form of the interaction is known as amplitude-coupling model and is such that $[H_{SR}, H_S] \ne 0$. In the amplitude-coupling model the decay of the off-diagonal elements of the system density matrix $\rho_S$ is accompanied by the exchange of energy and mode excitations between the system and the reservoir. This causes the fact that the diagonal elements of $\rho_S$, which count the occupation numbers of each state, evolve in time while the decoherence process occurs. In particular, within the amplitude-coupling model we will verify that the average occupation number of the system equilibrates with that of the reservoir. \\
We mention that in order to study quantum decoherence without exchange of energy and particle between the system and the reservoir, the interaction term should commute with the system hamiltonian $H_S$. For example, a Hamiltonian like 
\begin{equation}
    H_{SB} = a^\dagger a \sum \left( \kappa_j b_j + \kappa_j^* b_j^\dagger \right)
\end{equation}
known as phase-damping interaction, brings the loss of system coherences, so it allows to study entropy production by pure dephasing~\cite{Vidiella-Barranco:2016hnh,Coci:2025drb}; however the initial occupation numbers remain conserved. 
Let us briefly review how to derive the quantum master equation for the reduced density matrix of the system which is defined as
$ \rho(t) = Tr_R \left[ \rho_t  \right] $
taking the trace of the total density operator with respect to the large degrees of freedom of the reservoir. In standard techniques of OQS~\cite{carmichael1993quantum, BreuerPetruccione2002, weiss2008quantum} starting from the Liouville von-Neumann equation
\begin{equation}
    \frac{d \rho_t}{dt} = -i [H, \rho_t]
    \label{eq:Liouville}
\end{equation}
for $\rho_t$ after tracing over the environment degrees of freedom the first assumption to consider is the Born-Markov approximation which is valid for weak interaction. In the Born-Markov approximation the total density matrix can be factorized as
\begin{equation}
    \rho_t(t) \simeq \rho(t) \otimes \rho_R
\end{equation}
and the reservoir density matrix is taken to be that of a thermal heat bath
\begin{equation}
    \rho_R = \rho^{th}_R = \frac{e^{-\beta H_R}}{Z_R}
\end{equation}
where $\beta=1/T$ and $Z_R = Tr_R[e^{-\beta H_R}]$. Moving in the interaction picture and naming the bath operators
\begin{equation}
    B_I(t) = \sum_j \kappa_j(t) e^{-i \omega_j t} b_j \, , \, B^\dagger_I(t) = \sum_j \kappa^*_j(t) e^{+i \omega_j t} b_j^\dagger
\end{equation}
the trace over the reservoir degrees of freedom requires to compute the following correlation functions 
\begin{align}
    F(t) = \int_0^t \! dt' \langle B_I(t) B_I^\dagger(t') \rangle \\
    G(t) = \int_0^t \! dt' \langle B_I^\dagger(t) B_I(t') \rangle
\end{align}
and their corresponding cojugates, where $\langle \dots \rangle$ indicates the average over the statistical heat bath. Using the expression of the bath operators one gets~\cite{Brasil_2013}
\begin{align}
    F(t) &= \int_0^t dt' \! \sum_j \kappa_j \kappa^*_j \left( n_B(\omega_j, T) + 1 \right) e^{-i \omega_j(t-t')} \\
    G(t) &= \int_0^t dt' \! \sum_j \kappa_j \kappa^*_j  n_B(\omega_j, T)  e^{+i \omega_j(t-t')}
\end{align}
where the Bose-Einstein distribution $n_B(\omega,T)= \frac{1}{e^{\omega/T}-1}$ as function of the frequencies $\omega_j$ and the temperature $T$  appears. If the heat bath is static and the system-reservoir couplings do not depend on time the expression for $F(t)$ and $G(t)$ is usually integrated out by introducing a spectral density for a continuum of bath oscillators 
\begin{equation}
    J(\omega) = \sum_j |\kappa_j|^2 \delta(\omega - \omega_j)
    \label{eq:spectraldensity}
\end{equation}
and considering the Markovian limit for $t \rightarrow \infty$. In this case, the solution of time integral for the correlation functions is straightforward and back to the Schr\"odinger picture the master equation for the amplitude-coupling is~\cite{carmichael1993quantum}
\begin{eqnarray}
\frac{d\rho}{dt} &=& -i\omega^\prime_0[a^\dagger a,\rho]\nonumber\\
&&+
\frac{\gamma}{2}(1+\bar n)(2a\rho a^\dagger - a^\dagger a \rho - \rho a^\dagger a)
\nonumber \\
&&+\frac{\gamma}{2}\bar n(2a^\dagger \rho a  - a a^\dagger\rho - \rho aa^\dagger),
\label{eq:lindbGHJKL}
\end{eqnarray}
where $\bar{n} = n_B(\omega_0,T)$ is the Bose-Einstein distribution of the heat bath modes at the system frequency $\omega_0$ and $\gamma(T)$ is the real positive decay linewidth of the reservoir. In the first term of Eq.~\eqref{eq:lindbGHJKL} the frequency $\omega_0'$ accounts for the frequency renormalization due to the Lamb shift effect, but usually this correction is negligible so we will consider $\omega_0' \simeq \omega_0$. Eq.~\eqref{eq:lindbGHJKL} can be casted in the canonical Gorini-Kossakowski-Sudarshan-Lindblad~\cite{Lindblad:1975ef,Gorini:1975nb} with the two Lindblad operators
\begin{equation}
    L_1 = \sqrt{\gamma(\bar{n}+1)} \, a \, , \, L_2 = \sqrt{\gamma \bar{n}} \, a^\dagger
    \label{eq:lindope}
\end{equation}
so to write
\begin{equation}
    \frac{d \rho}{dt} = - i [H_0 , \rho] + \sum_{k=1}^2 \left( L_k \rho L_k^\dagger - \frac{1}{2} \lbrace L_k^\dagger L_k, \rho \rbrace \right)
\label{eq:lindblad}
\end{equation}
The resulting Lindblad Eq.~\eqref{eq:lindblad} captures the irreversible process of dissipation of the system within the thermal heat bath which also drives to quantum decoherence by preserving the complete positivity and the unitary trace of the density matrix. For atom or cavity modes coupled to radiation fields the Lindblad equation is obtained in the so-called optical regime which requires the following separation of time scales
\begin{equation}
    \tau_E \ll \tau_R \, , \, \tau_S \ll \tau_R 
\end{equation}
The first one indicates that bath correlation decays fast with characteristic time $\tau_E$ much smaller than the typical system relaxation time $\tau_R$ and allows to apply Born-Markov approximation both in the optical and quantum Brownian motion regime. The second expression is proper of the optical regime $ \gamma^{-1} = \tau_R \gg \tau_S = \omega^{-1}_0$ and indicates that rapidly oscillating terms at optical frequencies are neglected (rotating-wave-approximation).

Projecting Eq.~\eqref{eq:lindbGHJKL} on the Fock basis states $|m\rangle , |n\rangle$ one gets
\begin{eqnarray}
\dot\rho_{mn} &=& -i\omega_0(m-n)\rho_{mn} \nonumber \\
&+& \frac{\gamma}{2}(1+\bar n)(2 \sqrt{(m\!+\!1)(n\!+\!1)}\rho_{m+1,n+1}  - (m+n)\rho_{mn} )
\nonumber \\
&+& \frac{\gamma}{2}\bar n(2 \sqrt{mn} \rho_{m-1,n-1}  
- (m+1)\rho_{mn} - (n+1)\rho_{mn}) \nonumber \\
\label{eq:lindbladproj}
\end{eqnarray}

The expression Eq.~\eqref{eq:lindbladproj} corresponds to a set of coupled differential equations which can be solved numerically by simple Runge-Kutta methods, by discretizing the time variable $t_i = i \cdot \Delta t$ and truncating the Hilbert space corresponding to the maximum number of Fock states $N$. These parameters must be chosen so as to ensure that $\sum_ {n=0}^{N-1} \rho_{nn}(t_i) \simeq 1$ within reasonable numerical error.

\subsection{Extension with time dependent temperature}
In HICs the environment is represented by an expanding plasma of quarks and gluons (QGP) which evolves as a nearly perfect fluid. Including a time evolution of the environment is a quite complicated problem as it can alter the well defined time scales required to get a Markovian master equation in the Lindblad form.
On the one hand, in OQS the derivation of time-local master equations requires specific assumption which usually lead to the applicability of time-convolutioness projection operator technique on Nakagima-Zwanzig equations~\cite{BreuerPetruccione2002}. This topic is extensively treated in the literature and it is an ongoing topic of research in the OQS.
On the other hand, we want to study the effect of having an expanding reservoir by simply extending our model to include time dependence within the main parameters.
Therefore, we assume that we can factorize the time depedence within the coupling coefficients as follows
\begin{equation}
    \kappa_j(t) = \kappa_j \, f(t) \, , \, \kappa_j^*(t') = \kappa^*_j \, f(t')
\end{equation}
where now $\kappa_j$ are complex time independent quantities and $f(t)$ is a real positive function which models the system-reservoir coupling as function of time. Using again the spectral function $J(\omega)$ the correlation functions become
\begin{align}
    F(t) &= \int_0^t dt' f(t) f(t') \int_{0}^{\infty} d \omega J(\omega) (n_B(\omega, T(t')) + 1) e^{-i \omega(t-t')} \\
    G(t) &= \int_0^t dt' f(t) f(t') \int_{0}^{\infty} d \omega J(\omega) n_B(\omega, T(t')) e^{+i \omega(t-t')} 
    \label{eq:time-correlation}
\end{align}

Doing that, the possibility to exchange the frequency integral with the integral over time and sending the upper limit to $t \rightarrow \infty$ relies only on the temperature dependence $T(t')$ which appears in the Bose-Einstein distribution.  
Introducing an expanding environment implies that we have another time-scale included in the picture. We call this "hydrodynamical" time $\tau_{hydro}$ which controls the scale variation of the temperature and we consider valid the condition 
\begin{equation}
    \tau_E \ll \tau_{hydro} \le \tau_R \, ,
\end{equation}
which has the following meaning. On the one hand, the comparison of the time-scale at the extreme of the chain guarantees the Born-Markov approximation indicating that the emission of quantum modes during the system state transitions is immediately reabsorbed by the bath without affecting its properties. On the other hand, we assume that the expansion of the environment happens at sufficient large time that it does not affect the ability of restore thermalization within the time it takes the macroscopic temperature to suffer a sensitive variation due to the gradient expansion. 
Within this quasi-static approximation we can put in~\eqref{eq:time-correlation} $T(t^\prime) \simeq T(t)$ and perform first the time integration defining $\tau=t-t^\prime$ assuming Markov limit
\begin{equation}
    f(t) \int_0^{\infty} d \tau f(t-\tau) e^{-i(\omega-\omega_0) \tau} \simeq f^2(t) \int_0^{\infty} d\tau e^{-i(\omega-\omega_0) \tau} 
\end{equation}
where we have restored the oscillating term in Schr\"odinger picture and in the step we used again the condition $\tau \ll \tau_{hydro}$, since the system-reservoir coupling should change considerably only in a time interval where also the temperature of the reservoir changes. Finally, the last time integration produces the condition
\begin{equation}
\lim_{t \rightarrow \infty} \int_0^{t} d\tau e^{-i (\omega - \omega_0) \tau} = \pi \delta(\omega-\omega_0) - i \frac{PV}{\omega-\omega_0}
\end{equation}
where $PV$ is the principal value and which returns exactly the coefficients for the static case. These coefficients are multiplied by the time dependent function $f^2(t)$, This function is unknown, however, its effect can be reabsorbed within the coefficients which now can depend on time according to the variation of temperature.

As a result the quantum master equation remains of the form of Eq.~\eqref{eq:lindblad}, but in this case the Lindblad operators Eq.~\eqref{eq:lindope} have coefficients $\gamma$ and the density $\bar{n}$ which change according to the function $T(t)$.

\subsection{Model parameters}
The parameters of our model are the characteristic frequency of the system $\omega_0$ and the interaction coefficient $\gamma(T)$, which depends on the temperature of the bath.

To fix the value of $\omega_0$, we take inspiration from a recent approach that describes the production of charmonium and bottomonium using the Wigner density formalism~\cite{Song:2024got}. In particular, under the assumption that the $Q\bar{Q}$ interaction potential is that of a harmonic oscillator, the analytic Wigner density of the ground state in the phase-space is a Gaussian function characterized by a position width
\begin{equation}
    \sigma_r = \frac{1}{\sqrt{\mu \omega_0}}
    \label{eq:omega0}
\end{equation}
where $\mu = M/2$ is the reduced mass of the $Q\bar{Q}$ pair ($Q=c,b$). Then, in Ref.~\cite{Song:2024got} the width $\sigma_r$ is fixed to reproduce the experimental root-mean square radius $\langle r^2 \rangle$ of the real quarkonium wave function which for the ground state is related to the width by $\langle r^2 \rangle = 3 \sigma^2_r/2$. At this point, we can use the obtained value of $\sigma_r$, invert Eq.~\eqref{eq:omega0} to determine $\omega_0$. In particular, using the values of $\sigma_r=0.348$ fm for $J/\Psi$ 
reported in the Table 1 of Ref.~\cite{Song:2023zma}, with $M_c=1.4$ GeV 
we have $\omega_0 = 0.457$ GeV. 
We notice that under these conditions the binding energy is such that $E_B \simeq \omega_0 \ge T$, hence ensuring the validity of the optical regime.

In the damped harmonic oscillator model, the parameter $\gamma$ characterizes how strongly the system exchanges energy with the reservoir. 
A microscopic derivation of $\gamma$ demands knowledge of the heat bath spectral density defined in Eq.~\eqref{eq:spectraldensity} and introduced in the correlation functions so to get
\begin{equation}
    \gamma = 2 \pi J(\omega_0)
\end{equation}
where the spectral density is computed at the system frequency $\omega_0$. 
Hence, $\gamma$ encodes the rate at which the environment dissipates energy and destroys coherence driving the mean occupation number of the system towards equilibrium  $\langle a^\dagger a(t) \rangle \propto e^{-\gamma t} $. Then, it is clear that $\gamma$ can be identified as the inverse of the relaxation time $\tau_R = \gamma^{-1}$ which in the weak-coupling regime is much larger than the time $\tau_E$ at which energy is absorbed by the reservoir. In our model it is necessary that we provide a parametrization of the coefficient $\gamma$ as function of the temperature of the medium.
The dissociation time of tightly bound quarkonium states
in the QGP is the characteristic time required for the medium to break the heavy bound state through color screening, gluon dissociation, or inelastic scattering~\cite{Rapp:2009my}.
In this first study, we use a temperature-dependent dissociation rate for charmonium ground state ($J/\Psi$ particle) taken from the works~\cite{Grandchamp:2001pf,Grandchamp:2002wp}.

\section{Populations in a static heat bath}
\label{sec:III}

In this section, we discuss the results of our model for studying the dissociation of the ground-state of charmonium $J/\Psi$ 
in a static medium at fixed temperature $T$, whose value is typical of relativistic nuclear collisions. In particular, to validate our model, we compare the time evolution of the first diagonal element $\rho_{00}(t)$, i.e. the population of the ground state, with the standard classical evolution of the particle densities $n_i(t)$ obtained by solving the rate equations
\begin{equation}
\frac{d n_i(t)}{d t} = - \gamma(T) \left[ n_i(t) - n_i^{eq}(T,\omega_0) \right] \, .
\label{eq:rateeq}
\end{equation}

$\gamma(T)$ is the dissociation rate as a function of $T$, which has been used in the original study of charmonium suppression and regeneration~\cite{Grandchamp:2002wp,Rapp:2009my}, and which we employ in the dissipation part of the Lindblad equation Eq.~\eqref{eq:lindbGHJKL}.

\begin{figure}[h!]
    \centering
    \includegraphics[width=0.85\linewidth]{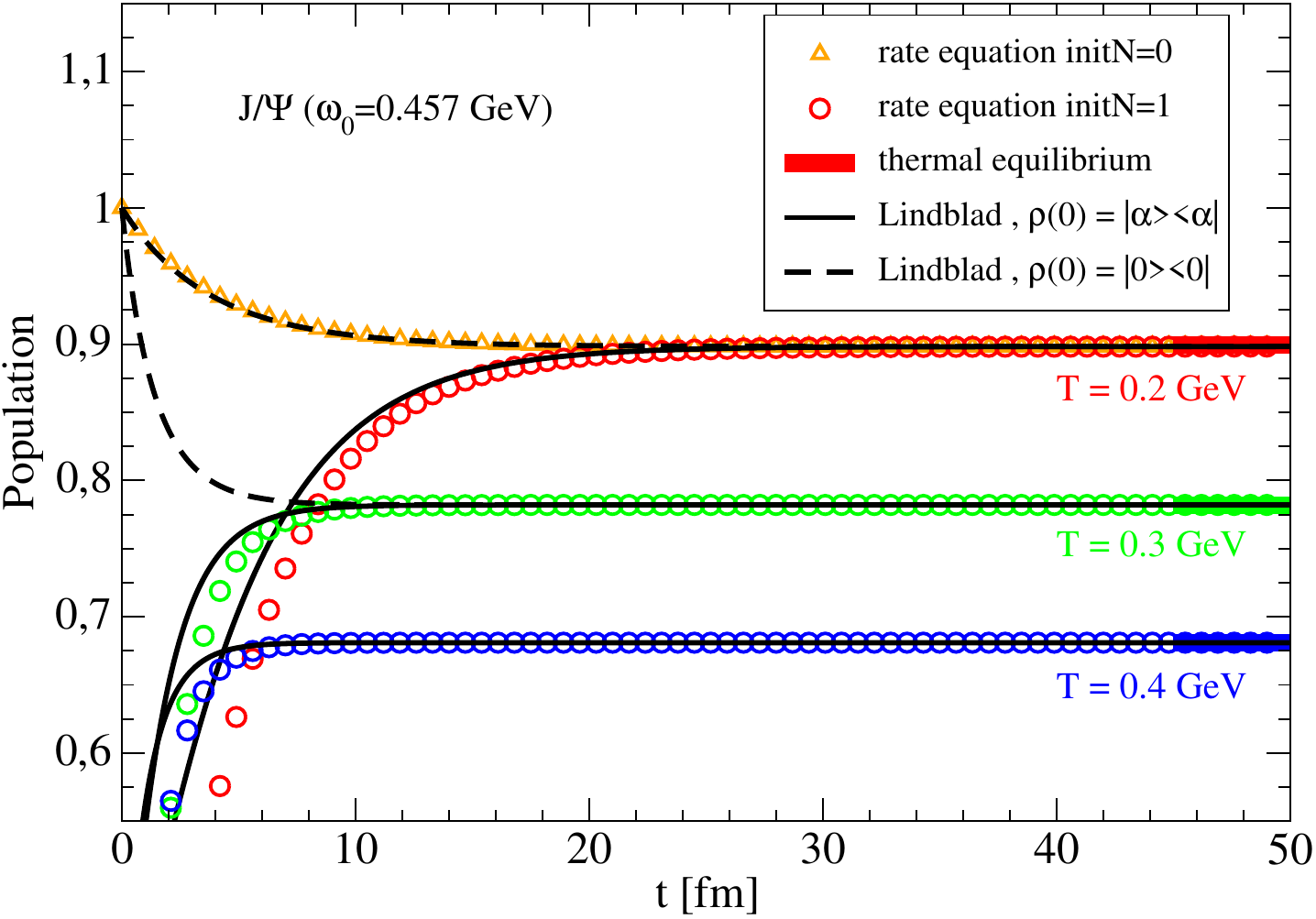}
    \caption{(color online): Population of $J/\Psi$ ground state as function of time in static heat bath at indicated temperature $T=0.2,0.3,0.4$ GeV. The black curves are the $\rho_{00}(t)$ elements obtained from the Lindblad equation for two initial conditions: ground state $|0 \rangle$ (dashed), coherent state $|\alpha \rangle$ (solid) with $\alpha=1$. Lindblad evolution is compared with results from rate equations. Both cases evolve asymptotically to thermal equilibrated values $n^{eq}(\omega_0,T)$ indicated with the thick colored strips and labeled with the $T$ values.}
    \label{fig:fig1}
\end{figure}

In Fig.~\ref{fig:fig1} we show the time evolution of charmonium ground state $J/\Psi$, which in our model corresponds to the occupation number of ground state of the H.O. system with frequency $\omega_0=0.457$ GeV.
The black curves represent the behavior of $\rho_{00}$ diagonal term obtained from the Lindblad Eq.~\eqref{eq:lindbGHJKL}. To this equation we associate an initial condition for the density matrix corresponding to a pure quantum state, such to have $Tr(\rho(0))=1$. In Fig.~\ref{fig:fig1} the dashed black line is the Lindblad evolution starting with initial fully occupied ground state, i.e. $\rho(0)=|0 \rangle \langle 0|$, while the solid black line represents the time evolution starting from a pure coherent state $\rho(0)=|\alpha\rangle\langle \alpha|$ with $\alpha=1$. We remind that a coherent state $|\alpha \rangle$ is a special quantum state of the H.O. which is defined as an eigenstate of the annihilation operator, i.e. $a |\alpha \rangle = \alpha | \alpha \rangle$ with complex parameter $\alpha$~\cite{Cahill:1969it,Walls:1985tm}. In the Fock basis a coherent state is given by the following expression
\begin{equation}
|\alpha\rangle = e^{-|\alpha|^2/2} \sum_n \frac{\alpha^n}{\sqrt{n!}} |n\rangle.
\label{eq:CSexpa_fock}
\end{equation}
Hence, the projector of a pure coherent state $|\alpha\rangle\langle \alpha|$ can be constructed using the Fock representation of the density operator with elements
\begin{equation}
\rho = \sum_{m,n} e^{-|\alpha|^2}
\frac{\alpha^m(\alpha^*)^n}{\sqrt{n!m!}} |m\rangle\langle n |,
\label{eq:RHOexpa_fock}
\end{equation}
where the sum over the Fock states in Eq.~\eqref{eq:RHOexpa_fock} is truncated to the maximum number $N$.
It is well known that a coherent state as a special role as it resembles the classical behavior of the quantum system, because it exhibits the minimal uncertainty with equal fluctuations in position and momentum. Therefore, it will be important for studying the behavior of quantum decoherence and the evolution of the Wigner function discussed in Sec.~\ref{sec:V}. 

Back to Fig.~\ref{fig:fig1} the black curves representing the Lindblad evolution of $\rho_{00}$ are confronted with the solution of the classical rate Eq.~\eqref{eq:rateeq} for three different bath temperatures: $T=0.2$, $0.3$ and $0.4$ GeV. To make possible such comparison we assume that the initial density $n_i$ in the studied volume is normalized to unity and we provide initial conditions which resemble those given for $\rho(0)$.
In particular, at $T=0.2$ GeV the orange triangles represent the case where $n_{J/\Psi}(0)=1$ showing perfect agreement with the Lindblad evolution with initial fully occupied ground state, i.e. $\rho(0)=|0\rangle \langle 0|$.
Still in Fig.~\ref{fig:fig1} the open circles represent the classical evolution where the first excited state is the only occupied initially. The colors of the circles indicate the three different values of $T=0.2$ GeV (red), $0.3$ GeV (green), $0.4$ GeV (blue).

We see that the quantum evolution from initial $\rho(0) =|\alpha \rangle \langle \alpha|$ with $\alpha=1$ follows quite well the trend of classical rate equations with initial density of first excited state set to unity. 
As we expect, both the quantum-Lindblad and the classical-rate solution of the occupation number of the ground state asymptotically reaches the thermal equilibrium value of the ground state which in the limit $E_B \simeq \omega_0 > T$ is given by the Boltzmann factor $e^{-E_B /T}$.

Our results indicate that independently of the provided initial conditions, the ground state thermalizes within a time-scale that decreases from 20 fm to 5 fm as the medium temperature increases from 0.2 GeV to 0.4 GeV.

We point out that the agreement between the quantum and the classical evolution is important to pose our question regarding the role of the decoherence process undergoing during dissipation and fluctuations which are driving the system towards equilibration with medium. In other words, the comparison between state populations legitimates the comparison of quantum master equation in optical regime with semi-classical approaches.

\section{Quantum decoherence}
\label{sec:IV}
Now we focus on the quantum decoherence process, by analyzing the time evolution of the off-diagonal terms of the system density matrix, known as coherences. From Eq.~\eqref{eq:lindbladproj} we can infer the analytic behavior in the large-time limit $\gamma t \gg 1$ of high-order coherences. Taking $m=0$ and $n \gg 1$ the right-hand side of the Lindblad Eq.~\eqref{eq:lindbladproj} can be approximated to
\begin{equation}
    \dot\rho_{0n} \approx \left[ i \omega_0 n - \frac{\gamma}{2} \left( n(1 + 2\bar{n}) + 2 \bar{n} \right) \right] \rho_{0n} \, .
    \label{eq:coherencem0}
\end{equation}
Then, retaining the larger terms in the square brackets which are proportional to $n \gg 1$, the approximate solution of Eq.~\eqref{eq:coherencem0} is
\begin{equation}
    \rho_{0n}(t) = \rho_{0n}(0) e^{-a_n t}
\label{eq:rhom0}
\end{equation}
with the coefficient at exponential defined as
\begin{equation}
a_n = -i n \omega_0 + \frac{\gamma}{2} n \left( 1 + 2 \bar{n} \right)
\label{eq:am0}
\end{equation}
As function of time the coherence elements $\rho_{0n}$ follow an oscillating behavior with frequency $n \omega_0$ proportional to the intrinsic frequency of the system $\omega_0$. These oscillations are also damped by the real part of the $a_{n}$ coefficient which is proportional to the dissociation rate $\gamma$ and the mean occupation number of the bath $\bar{n} = n_B(\omega_0,T)$. Hence, the stronger the system-reservoir coupling, the faster the decay of the off-diagonal elements $\rho_{0n} \rightarrow 0$, which clearly indicates that asymptotically full decoherence is achieved and the density matrix becomes that of an incoherent mixture of states 
\begin{equation}
    \rho^{\infty} = \sum_{n} \rho^{th}_{nn} | n \rangle \langle n |
\end{equation}
where the diagonal elements correspond to the occupation numbers $\rho_{nn}^{th}$ in thermal equilibrium with the bath.
We compare the result for the matrix elements $\rho_{0n}$ obtained by numerically solving the Lindblad equation Eq.~\eqref{eq:lindbGHJKL}, with the approximate values given by Eq.~\eqref{eq:coherencem0}. 
To study the decoherence process, the initial density matrix must be such that its off-diagonal elements $\rho_{nm}$ are non-zero; in this case, it is not possible to set $\rho(0)$ from a pure state of the Fock basis which is fully occupied at initial time. However, we can choose as a simple initial condition the one given from a coherent state $|\alpha \rangle$ whose matrix elements can be read from Eq.~\eqref{eq:RHOexpa_fock} with parameter $\alpha$ set to 1.

\begin{figure}[h!]
    \centering
    \includegraphics[width=\linewidth]{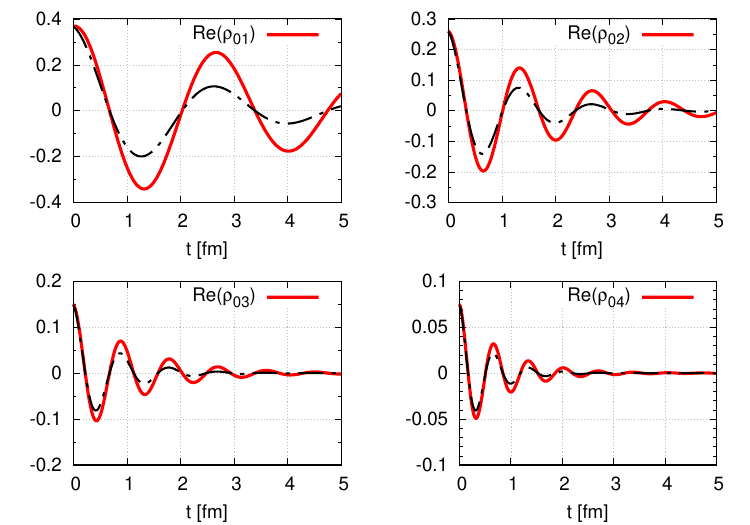}
    \caption{Time evolution of the real part of coherence elements $\rho_{0n}$ with $n=1,2,3,4$ (from top left to bottom right). The thick solid red line is the result obtained by numerically solving the Lindblad equation with initial density matrix $\rho(0)=|\alpha \rangle\langle \alpha|$ with $\alpha=1$, frequency $\omega_0=0.457$ GeV and bath temperature $T=0.3$ GeV. The dash-dotted black line corresponds to the analytic trend given by Eq.~\eqref{eq:coherencem0} with $a_n$ coefficient from Eq.~\eqref{eq:am0} assuming $n \gg 1$.}
    \label{fig:ReCoherenceT300}
\end{figure}

\begin{figure}[h!]
    \centering
    \includegraphics[width=\linewidth]{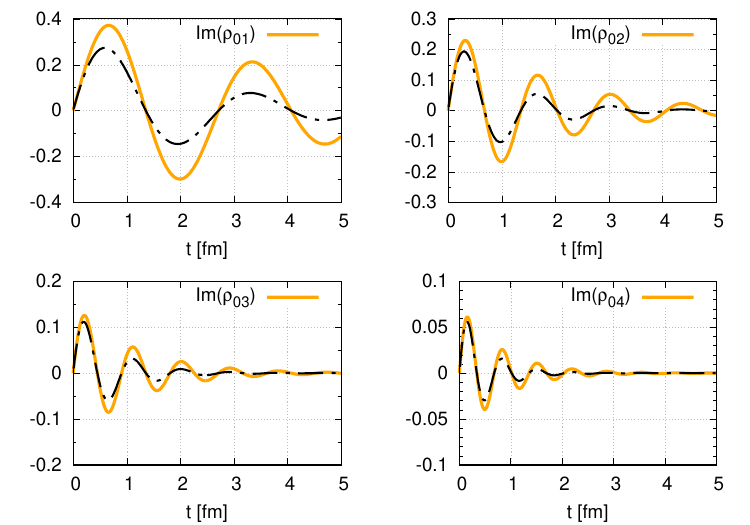}
    \caption{Time evolution of imaginary part of coherence elements $\rho_{0n}$ with $n=1,2,3,4$ (from top left to bottom right). The thick solid orange line is the numerical result from the Lindblad equation and the dash-dotted black line is the approximate analytic behavior in Eq.~\eqref{eq:coherencem0}. The model parameters are the same reported in the caption of Fig.~\ref{fig:ReCoherenceT300}.}
    \label{fig:ImCoherenceT300}
\end{figure}

In Fig.~\ref{fig:ReCoherenceT300} and Fig.~\ref{fig:ImCoherenceT300} we show the time evolution of the real (thick solid red line) and imaginary part (thick solid orange line) of the coherence elements $\rho_{0n}(t)$ with $n=1,2,3,4$  
for system frequency $\omega_0=0.457$ GeV and heat bath temperature $T=0.3$ GeV. The dash-dotted black line in both figures is the approximate analytic expression given by Eq.~\eqref{eq:coherencem0}. On the one hand the numerical results show the expected oscillating behavior of each $\rho_{0n}$ element with frequency $n \omega_0$. On the other hand as the number $n$ becomes larger, the decay of these oscillations is faster and agrees better with the damping effect provided by the real part of the coefficient $a_n$ in Eq.~\eqref{eq:am0}. Only for $n=1$ and $n=2$ the discrepancy between the numerical result and the approximate analytic solution is relevant passing from about $40 \%$ to $25 \%$ at the first minimum of both curves, while such discrepancy becomes negligible at $n=4$ indicating that the assumption $n \gg 1$ already works well.

We point out that a correct solution of the reduced density matrix $\rho(t)$ can be obtained by computing the correlation functions of the reservoir which in the end would provide the exact analytic behavior of the coherence elements. However, this requires the computation of exponential operators containing non-trivial commutators like $[H_{SR}(t),H_{SR}(t_1)]$, $[H_{SR}(t),[H_{SR}(t_1),H_{SR}(t_2)]]$. This can be performed iteratively or by means of the Magnus expansion~\cite{Magnus:1954zz} in order derive the so-called decoherence function $\rho_{mn}(t) = \rho_{mn}(0) D_{mn}(t)$. In particular, in the Markovian regime this function should be of the form $D_{mn}(t) = e^{-\Gamma_{mn}t}$ with $m \ne n$.
The correct evaluation of $D_{mn}(t)$ for quarkonium dissociation in OQS approach will be subject of a future study, while here we want to provide an estimate of the decoherence time as function of the temperature of the heat bath.

\begin{figure}[h!]
    \centering
    \includegraphics[width=\linewidth]{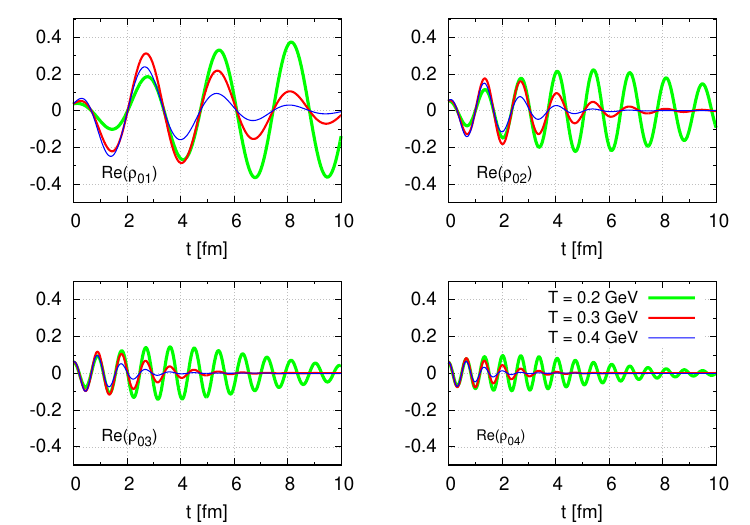}
    \caption{(color online): Time evolution of the real part of coherence elements for three different heat bath temperatures $T=0.2-0.4$ GeV. The frequency of the system is set to $\omega_0=0.457$ GeV.}
    \label{fig:REcoherenceTemp}
\end{figure}

In Fig.~\ref{fig:REcoherenceTemp} we show the time evolution of the same four off-diagonal elements $\rho_{0n}(t)$ with $n=1,2,3,4$ for $J/\Psi$ system initialized as a coherent state with $\alpha=1$ at three different temperatures. In particular, to the case at $T=0.3$ GeV (solid red curve also in Fig.~\ref{fig:ReCoherenceT300}) we add  $T=0.2$ GeV (solid green curve) and $T=0.4$ GeV (thin blue curve) to cover a wide range of temperatures explored by the QGP in HICs. As expected, we observe that with increasing $T$ the system-coupling strength increases leading to a faster decoherence process which removes the initial correlations of the coherent state superposition in about $5$ fm.

\section{Evolution of the Wigner distribution}
\label{sec:V}

\subsection{Fluctuation and dissipation}

The Lindblad Eq.~\eqref{eq:lindbGHJKL} derived in the optical regime with amplitude-coupling interaction can be reformulated by projecting it onto a basis other than the Fock basis, like for example the coherent-state basis~\cite{carmichael1993quantum,ScullyZubairy1997,weiss2008quantum}. In this representation, it is known that the time evolution of the density matrix is mapped onto a Fokker–Planck equation~\cite{risken1989fpe} for quasi-probability distribution. In quantum optics, this formulation provides a physically transparent description of decoherence: the suppression of off-diagonal elements of the density matrix is reflected in the progressive damping of the negative oscillatory fringes of the quasi-probability distribution, ultimately leading to a classical-like behavior. In particular, the Wigner function~\cite{Wigner:1932eb,Hillery:1983ms} offers a phase-space $(x,p)$ representation of this process, where decoherence manifests as a smoothing of the distribution and the gradual disappearance of its negative regions.
In this section, we first introduce the Wigner distribution through the density matrix of the quarkonium system. Then, we study its change in time resulting from the quantum Lindblad evolution by interaction within the bosonic heat bath at constant temperature.

The Wigner distribution at time $t$ is defined as the Weyl transform of the density matrix $\rho(t)$~\cite{Case:2008ped}~\footnote{In this section the coordinate $x$ and $p$ are meant as adimensional quantities: $x \rightarrow x \cdot \beta$, $p \rightarrow p / \beta$ where $x$ is in $fm$ and $p$ in $fm^{-1}$ while the coefficient $\beta=\sqrt{\mu \omega_0}$ is in $fm^{-1}$.}
\begin{equation}
    W(x,p,t) = \frac{1}{\pi} \int_{-\infty}^{\infty} \! dy \, e^{2ipy} \langle x-y | \rho(t) | x+y \rangle \, .
\end{equation}
We introduce the representation of the reduced density matrix $\rho(t)$ in terms of the Fock basis
\begin{equation}
    \rho(t) = \sum_{m,n} \rho_{mn}(t) | m \rangle \langle n|
\end{equation}
where the coefficients $\rho_{mn}(t)$ are those obtained as solution of the Lindblad Eq.~\eqref{eq:lindbGHJKL}. Doing this we get
\begin{eqnarray}
    W(x,p,t) &=& \frac{1}{\pi} \int_{-\infty}^{+\infty} \! dy \, e^{2ipy} \, \sum_{m,n} \rho_{mn}(t) \langle x-y| m \rangle \langle n | x+y \rangle \nonumber \\
    &=& \sum_{m,n} \rho_{mn}(t) W_{mn}(x,p)
    \label{eq:Wignerf}
\end{eqnarray}
where the coefficients $W_{mn}(x,p)$ are related to the Weyl transformation of the H.O. eigenfunctions. For this quantities we can use well known analytical formulas
for the diagonal terms $m=n$
\begin{equation}
    W_{nn}(x,p) = \frac{(-1)^n}{\pi} e^{-r^2} L_n^{(0)}(2r^2)
    \label{eq:WnnHO}
\end{equation}
and for off-diagonal terms $m > n$
\begin{equation}
    W_{mn}(x,p) = \frac{(-1)^n}{\pi}  \sqrt{\frac{n!}{m!}} (2\alpha)^{m-n} e^{-2|\alpha|^2} L_n^{(m-n)}(4|\alpha|^2)
    \label{eq:WmnHO}
\end{equation}
($W_{mn} = W_{nm}^*$ for $m < n$). In Eqs.~\eqref{eq:WnnHO} and \eqref{eq:WmnHO} $\alpha = \frac{x+ip}{\sqrt{2}}$, $r^2 =2|\alpha|^2= x^2 + p^2$ and $L_n^{(m-n)}$ are the generalized Laguerre polynomials. Eq.~\eqref{eq:Wignerf} allows us to compute the Wigner function during the Lindblad evolution of the density matrix $\rho(t)$. Numerically, we divide the phase space $(x,p)$ into a grid in coordinate $\Delta x$ and momentum $\Delta p$ and compute the Wigner function at each point according to Eq.~\eqref{eq:Wignerf}. As we ensure the trace of the density matrix is preserved 
during the time evolution, we check also that the Wigner function is correctly normalized within the numerical errors, namely
\begin{equation}
  \int_{-\infty}^{+\infty} \! dx \int_{-\infty}^{+\infty} \! dp \, W(x,p)
  = Tr{[\rho(t)]} = 1 \, .
  \label{eq:WignerNorm}
\end{equation}

\begin{figure}[h!]
    \centering
    \includegraphics[width=0.3\textwidth]{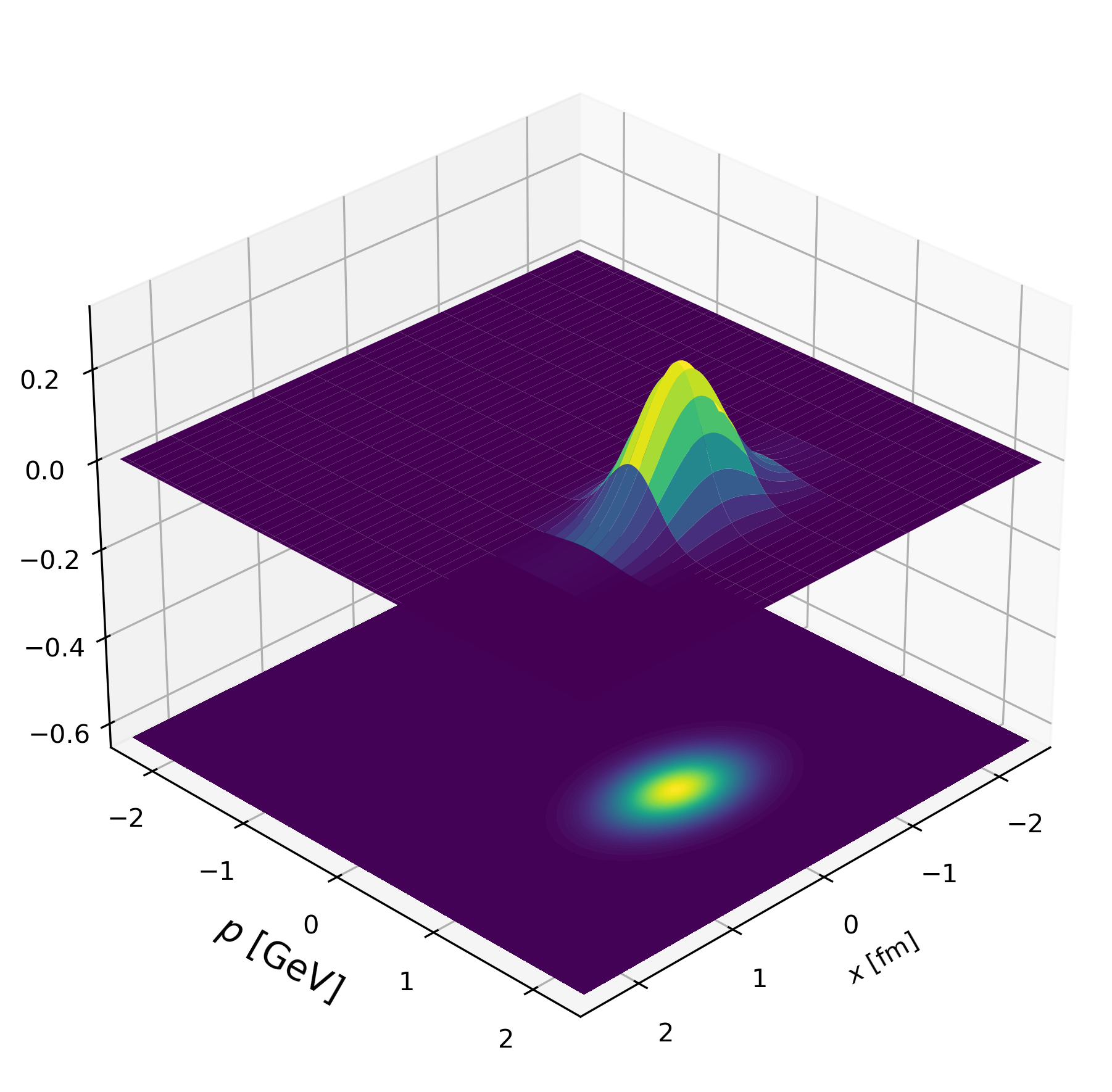}
    \caption{Contour plot of Wigner distribution for initial coherent state $|\alpha \rangle$ with $\alpha=2$.}
    \label{fig:wignerinitial}
\end{figure}

\begin{figure*}[t!]
    \centering
    \includegraphics[width=\linewidth]{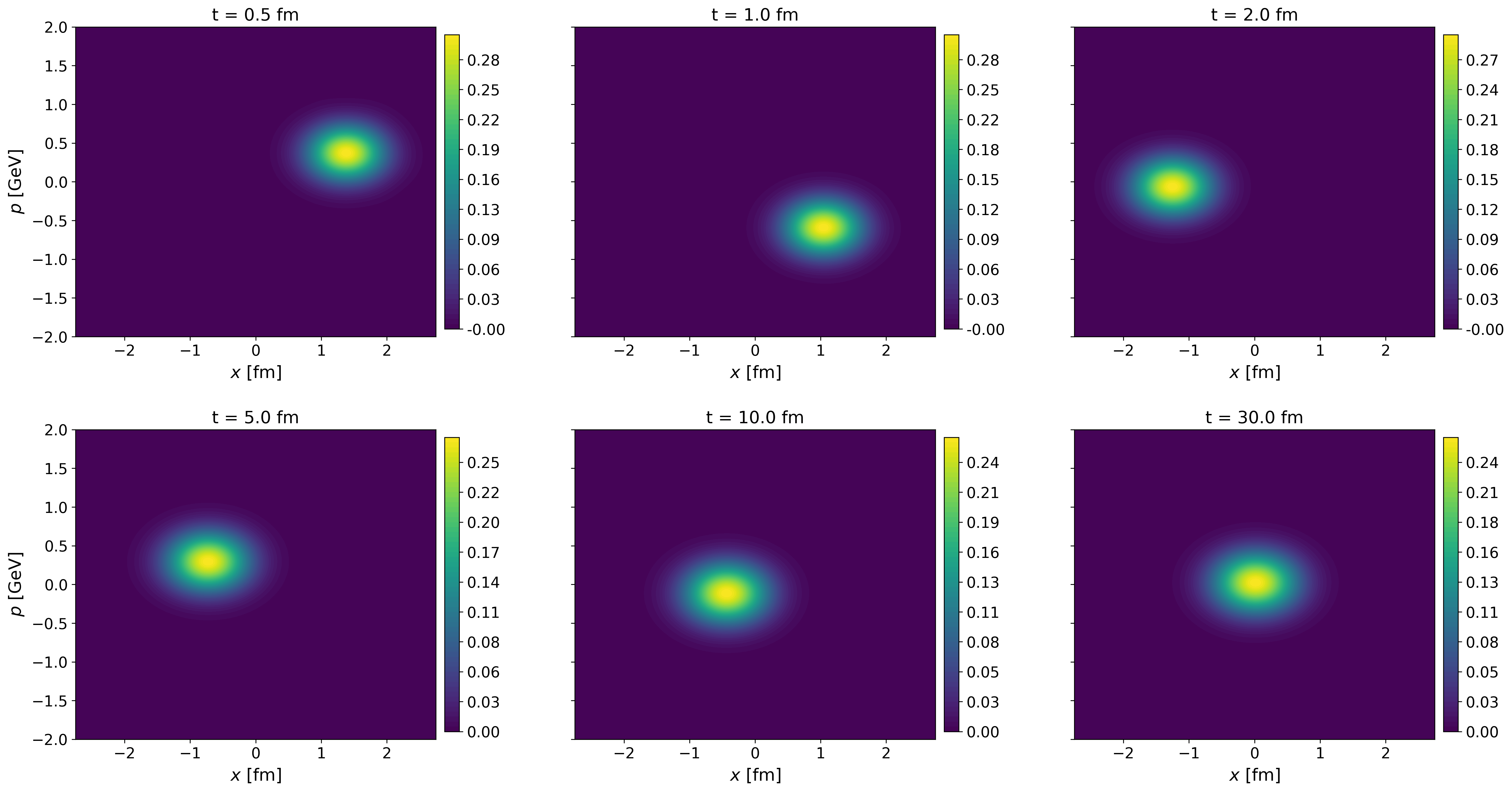}
    \caption{Time evolution of Wigner function $W(x,p)$ at fixed medium bath $T=0.2$ GeV starting from initial coherent state $|\alpha\rangle$ with $\alpha=2$. At final time $t=30$ fm the Wigner function reaches the asymptotic distribution from analytic expression.}
    \label{fig:wignertimeevo}
\end{figure*}

Moreover, we can use the expression of the density matrix in the large-time limit $\rho^{\infty}$, which is diagonal due to the quantum decoherence, to derive the analytical expression of the asymptotic Wigner distribution 
\begin{align}
W(x,p,t\rightarrow \infty) &= \sum_{n=0}^{\infty} \rho^{\infty}_{nn} W_{nn}(x,p) 
\label{eq:asymtoticWigner}
\\ 
&= \sum_{n=0}^{\infty} e^{-(n+\frac{1}{2})\frac{\omega_0}{T}} \, 2 \sinh{\left(\frac{\omega_0}{2T}\right)} W_{nn}(x,p) \nonumber
\end{align}
which is correctly normalized to unity and which we can confront to the result coming from the numerical solution of the Lindblad Eq.~\eqref{eq:lindbGHJKL} at large time $\gamma t \gg 1$. 
We study the evolution of the Wigner function of the quarkonium system interacting with a reservoir at constant temperature starting from different initial conditions.

In Fig.~\ref{fig:wignerinitial} we show the Wigner function and its contour profile in the phase-space associated to an initial coherent state $|\alpha \rangle$ from Eq.~\eqref{eq:CSexpa_fock} with parameter $\alpha=2$. 
We let the density matrix of the system evolve according to the Lindblad master equation Eq.~\eqref{eq:lindblad} with system frequency $\omega_0=0.457$ GeV and dissociation rate $\gamma$ computed for a thermal heat bath at fixed temperature $T=200$ MeV. In Fig.~\ref{fig:wignertimeevo} we show the three-dimensional profile and the contour plot of the associated Wigner function in the $(x,p)$ plane at different time snapshots from $t=0.5$ fm (upper left plot) to $t=30$ fm (bottom right plot).

\begin{figure*}
    \centering
    \includegraphics[width=\linewidth]{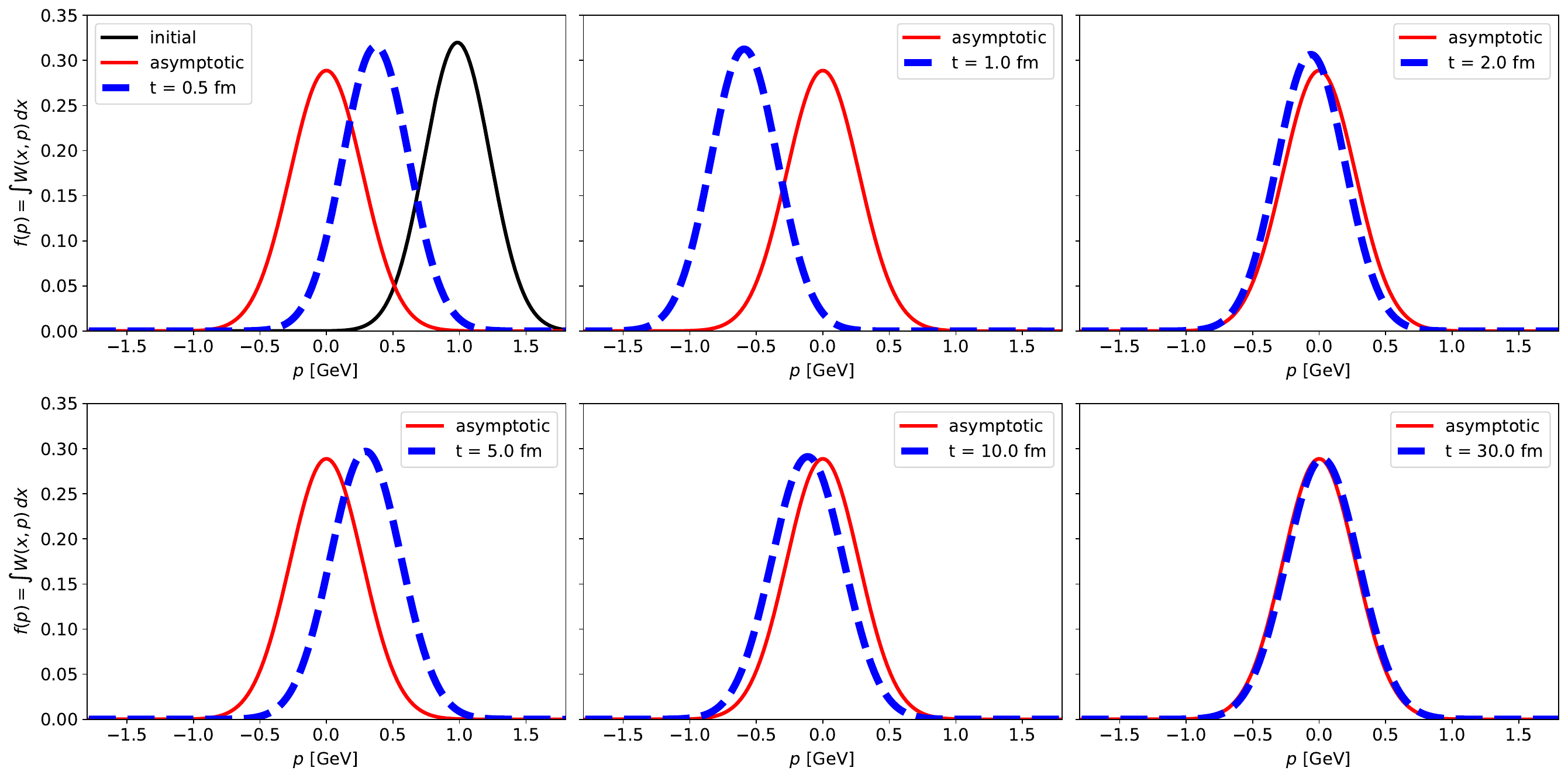}
    \caption{Time evolution of momentum distribution $f(p)$ at fixed medium bath $T=0.2$ GeV starting from initial coherent state $|\alpha\rangle$ with $\alpha=2$ (black line). The distribution $f(p)$ at different times is shown with the thick dashed blue line. }
    \label{fig:pdistrtimeevo}
\end{figure*}

In phase space, the dissipative and decoherence evolution of the initial coherent state is characterized by damped motion in the $x$ and $p$ variables, accompanied by diffusion that drives the Wigner function towards an incoherent mixture of localized (gaussian) wave packets.  
This asymptotic expression corresponds to the steady-state solution of the Lindblad equation Eq.~\eqref{eq:asymtoticWigner}, which in our case is represented by a mixture of H.O. eigenfunctions with thermal weights and in which off-diagonal terms are suppressed in the Fock basis.
As stated in Ref.~\cite{Zurek:1991vd} this steady solution can be considered a classical probability distribution.
Numerically, we verify that the Wigner function at $t=30$ fm is already consistent with the asymptotic expression~\eqref{eq:asymtoticWigner}. 
The dissipative phenomenon can be analyzed by means of the momentum ($p$) distribution by integrating the Wigner function in the coordinate $x$ and defining the momentum distribution as
\begin{equation}
    f(p) = \int \! dx W(x,p)
    \label{eq:fpdistr}
\end{equation}

According to Eq.~\eqref{eq:WignerNorm} the integral of the $f(p)$ distribution also needs to be 1.
In Fig.~\ref{fig:pdistrtimeevo} starting from the initial distribution (solid black line on the upper right plot) the evolution of $f(p)$ defined in Eq.~\eqref{eq:fpdistr} is shown with a dashed blue line at
different time snapshots from $t=0.5$ fm (upper left plot) to final $t=30$ fm (bottom right plot). In particular,
the distribution $f(p)$ at final time $t=30$ matches the distribution obtained from the asymptotic expression of $W(x,p,t\rightarrow \infty)$ shown with a solid red line in all the plots. One observes that the oscillations of the $f(p)$ curve reduce in amplitude due to the friction effect and that the peak lowers, thus indicating a broadening caused by diffusion. 
This proves that the function $f(p)$ follows the evolution of a Fokker-Planck equation.
The average relative momentum of the $Q\bar{Q}$ pair is defined as
\begin{equation}
    \langle p \rangle = \int_{-\infty}^{+\infty} \! dp \, p \, f(p)
\end{equation}

\begin{figure}[h!]
    \centering
    \includegraphics[width=\linewidth]{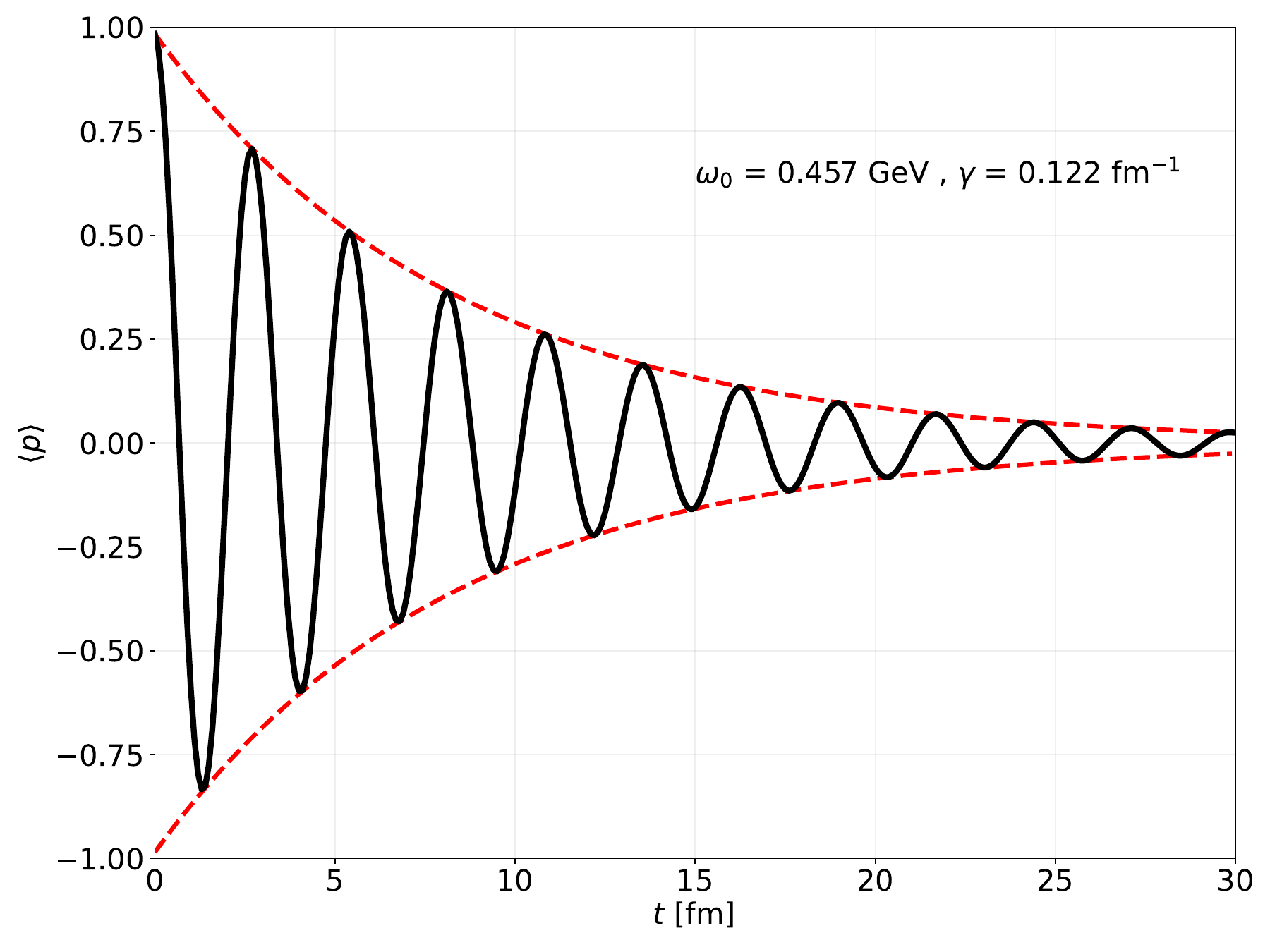}
    \caption{Average momentum $\langle p \rangle$ as function of time extrapolated from the damped oscillation of the $p$-distribution Eq.~\eqref{eq:fpdistr} (solid black line). The parameters $\omega_0$ and $\gamma$ indicate the frequency of oscillation and the damping rate obtained from the exponential fit (dashed red line).}
    \label{fig:paverage}
\end{figure}

and its behavior obtained by the numerical solution of the master equation is shown with the solid black line in Fig.~\ref{fig:paverage} for $J/\Psi$ system ($\omega_0=0.457$ GeV) interacting with reservoir at fixed temperature $T=0.2$ GeV. We observe that the oscillating behavior around the average zero relative momentum of the asymptotic gaussian distribution happens with a reducing amplitude due to the drag effect. In Fig.~\ref{fig:paverage} the dashed red lines correspond to a fit of the damped oscillations in the relative momentum which follow an exponential trend $\langle p \rangle(t) = p_0 e^{- \frac{\gamma}{2} t}$ where $p_0$ is the initial relative momentum and $\gamma$ is the only parameter which by fitting the exponential trend we find to be equivalent to the input dissociation rate from the TAMU model~\cite{Grandchamp:2002wp,Rapp:2009my}. This is the typical behavior for the evolution of $\langle p \rangle$ in the Fokker-Planck equation with a relaxation time $\tau_R = 2 \gamma^{-1}$.

\begin{figure}[h!]
    \centering
    \includegraphics[width=\linewidth]{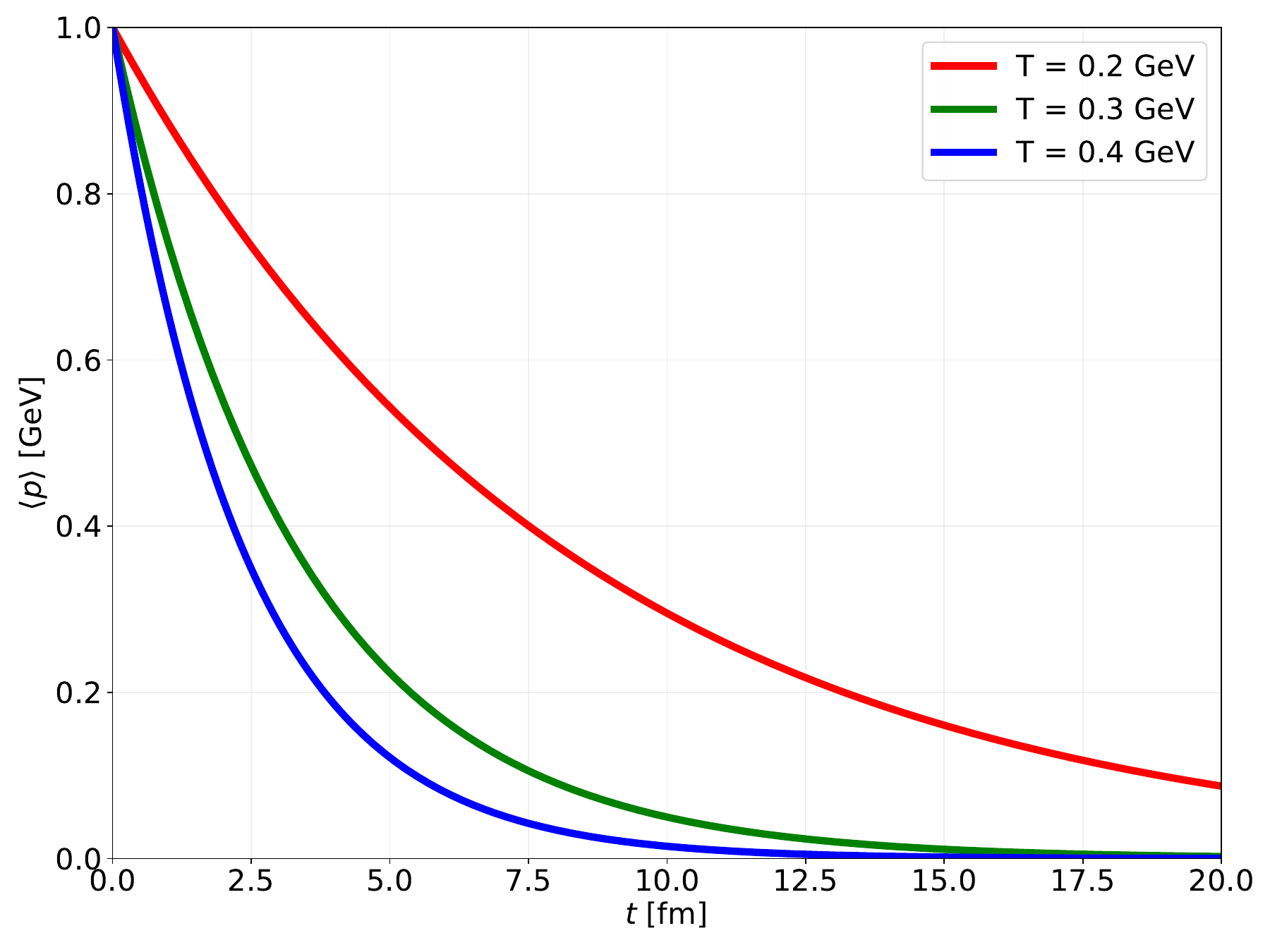}
    \caption{(Color online) Average relative momentum $\langle p \rangle$ as function of evolution time $t$ obtained from exponential fit of the damped oscillations for $J/\Psi$ system ($\omega_0=0.457$ GeV) at different heat bath temperature.}
    \label{fig:averagemomentum}
\end{figure}

To conclude, in Fig.~\ref{fig:averagemomentum} we show the obtained exponential damping of average relative momentum for $J/\Psi$ system ($\omega_0=0.457$ GeV) evolving in time via amplitude-coupling Lindblad master equation within a static heat bath at fixed temperature T=0.2, 0.3 and 0.4 GeV.

\subsection{Decoherence and classicalization}
 In the previous section we studied the time evolution of the Wigner distribution starting with initial condition of a coherent state $\rho(0)=|\alpha \rangle \langle \alpha |$. As shown in Fig.~\ref{fig:wignerinitial} and ~\eqref{fig:wignertimeevo} the Wigner function is always positive and the decoherence process is hidden in the fluctuation-dissipation effect on the distribution of the relative momentum $p$ of the $Q \bar{Q}$ pair.  However, as extensively discussed in the literature~\cite{Zurek:1991vd,Zurek:2003zz,Schlosshauer:2019ewh} decoherence is responsible also for the transition from quantum to classical behavior of the system as quantum coherence is leaked within the environment degrees of freedom. This corresponds to the suppression of all negative fringes in the Wigner distribution which are produced by non-zero quantum coherences. 
 Therefore, to study this effect we initialize the system using a "cat" state which is defined as the quantum superposition of two coherent states with opposite phases~\cite{Yurke:1986zz,Saito:2003hv}. In particular, we use the even cat state 
 \begin{equation}
     | \text{cat} \rangle_e = \frac{1}{\mathcal{N}} \left( | \alpha \rangle + |-\alpha \rangle \right)
     \label{eq:catstate} \,
 \end{equation}
where $\alpha$ is the usual complex number associated to coherent state, while $\mathcal{N}$ is the normalization factor. 

\begin{figure}[h!]
    \centering
    \includegraphics[width=0.3\textwidth]{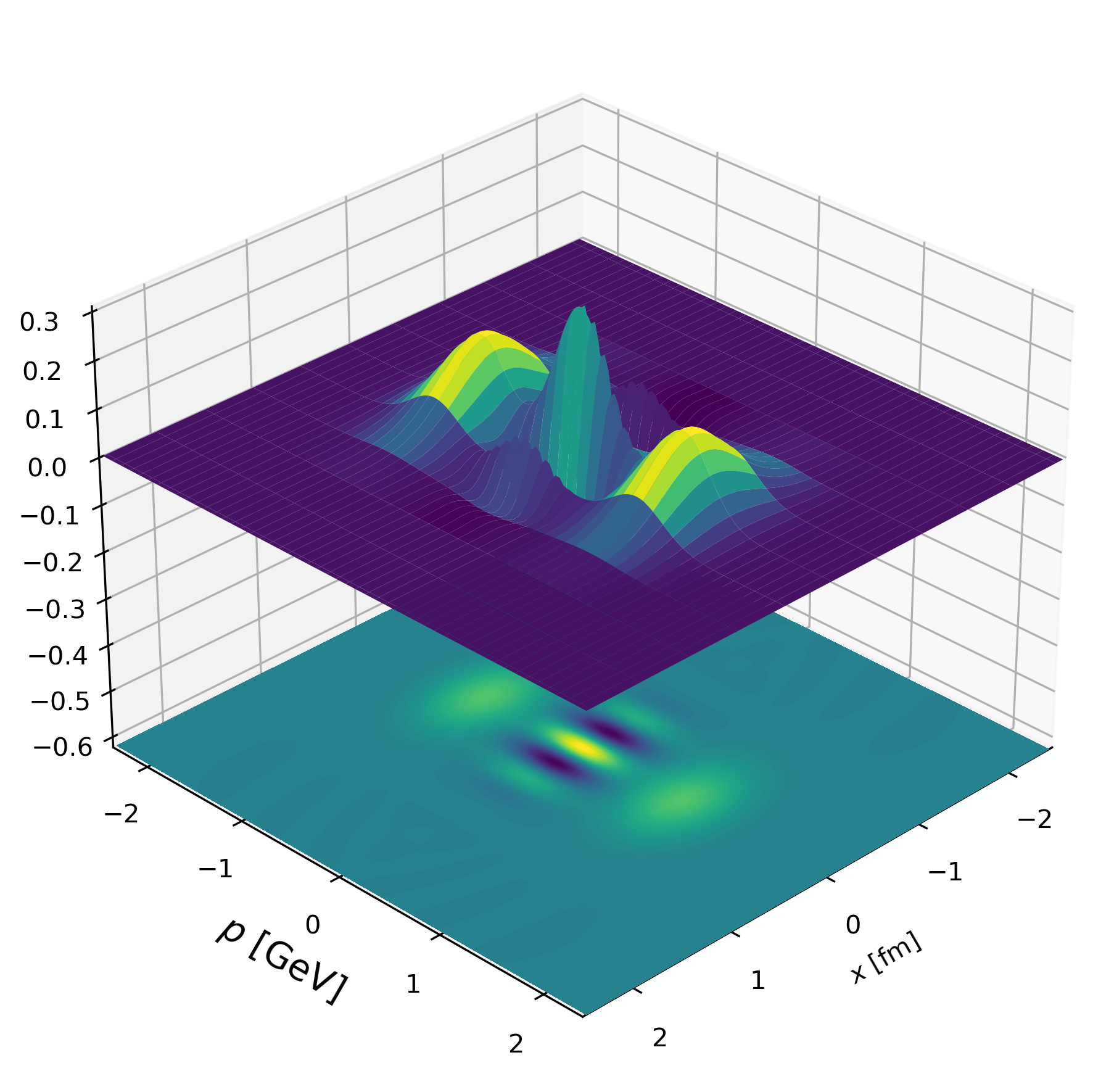}
    \caption{Contour plot of initial cat state Eq.~\eqref{eq:catstate} with parameter $\alpha=2$.}
    \label{fig:wignerinitcat}
\end{figure}

The corresponding Wigner function of the state Eq.~\eqref{eq:catstate} with $\alpha=2$ is shown in Fig.~\ref{fig:wignerinitcat}. We notice two positive peaks located at $p=\pm 1$ GeV. In this regard, we may think to simulate an initial $Q\bar{Q}$ pair produced with back-to-back momentum $p_Q = - p_{\bar{Q}} = p/2$ by hard scattering process at same location $x=0$, where quantum correlations are indicated by the interference terms in the central region.

\begin{figure*}[t!]
    \includegraphics[width=\linewidth]{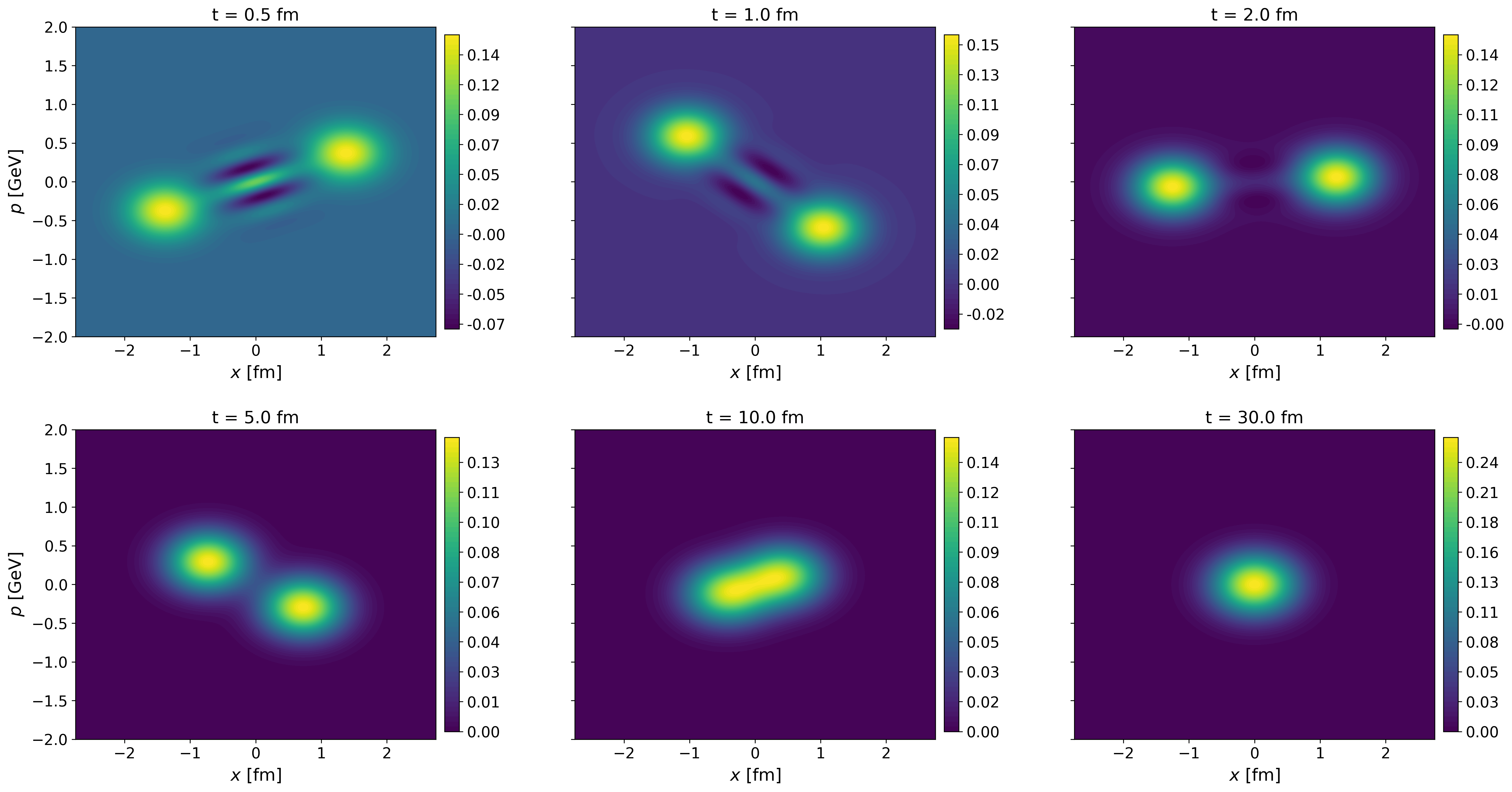}
    \caption{Contour plots of Wigner function $W(x,p)$ in phase-space grid at different times from $t=0.5$ fm (top left) to final $t=30$ fm (bottom right) constructed from density matrix $\rho_{Q \bar{Q}}(t)$ starting with initial cat state Eq.~\eqref{eq:catstate} with $\alpha=2$ and evolving via Lindblad equation with static heat bath at fixed $T=0.2$ GeV.}
    \label{fig:wignertimeevocat}
\end{figure*}

The time evolution of the Wigner function from Lindblad Eq.~\eqref{eq:lindbGHJKL}with initial "cat" state Eq.~\eqref{eq:catstate} for the case of static heat bath at $T=0.2$ GeV is shown in the successive plots of Fig.~\ref{fig:wignertimeevocat}. As it is clearly visible, the initial interference structure as well as the negative fringes are suppressed as the quantum coherences are washed out within a time of about $\tau \simeq \gamma^{-1} = 10$ fm for temperature of $T=0.2$ GeV.
Although this pure decoherence effect in the Markovian regime is well known in various areas of quantum optics, we believe that it may also play a role during the hadronization process. 
On the one hand, $Q\bar{Q}$  pairs produced at mid-rapidity experience the full space-time evolution of the fireball, interacting with the QGP and approaching almost thermal equilibrium, which ultimately leads to complete loss of quantum coherence. This, in turn, justifies the use of hadronization models based on coalescence with localized Wigner functions, typically parameterized as gaussians, which provide a classical probabilistic description of hadron formation.
On the other hand,  pairs produced at the phase-space border of the fireball may hadronize at earlier times without achieving full decoherence, hence  conserving part of their initial correlations. 
Therefore, this work can provide the basis for future studies about the role of decoherence effect within the Wigner function evolution in the heavy flavor hadronization process based on coalescence model~\cite{Plumari:2017ntm,Minissale:2023dct,Minissale:2024gxx}.

\section{Decoherence in evolving medium}
\label{sec:VI}
We now move to the study of quantum decoherence of heavy-quark bound states interacting with an evolving environment. To this end, we employ the Lindblad master equation for the system–environment interaction, where the coupling exhibits an explicit time dependence, as described in Sec.~IIB.
Our primary goal is to investigate the role of quantum decoherence in the suppression of quarkonium states produced in relativistic heavy-ion collisions. To this end, the thermal environment interacting with the heavy-quark bound state should mimic the dynamical evolution of the QGP created in the central region of nucleus--nucleus collisions at RHIC and LHC energies. As a first step, we consider a simplified one-dimensional expansion of the medium, in which the temperature evolves with time according to a power-law behavior
\begin{equation}
    T(t)=T_0 \left ( \frac{\tau_0}{t} \right)^{c_s^2}
    \label{eq:EvoTemp}
\end{equation}
where $T_0$ is the temperature at the initial time $\tau_0$ of fireball creation.
This setup should allow us, firstly, to isolate and quantify the impact of the time-dependent medium evolution on the decoherence dynamics of quarkonium-like states. Moreover, it is employed already in other models which study quarkonium suppression within the open quantum systems framework~\cite{Brambilla:2016wgg}.
In Eq.~\eqref{eq:EvoTemp} the parameter $T_0$ is fixed within the typical range of initial temperature of the fireball reached at RHIC and LHC collisions $T_0 \simeq 0.3-0.5$ GeV, while we set $\tau_0 = 0.6$ fm in agreement with the value of proper time inspired by Bjorken hydrodynamics~\cite{Bjorken:1982qr}. We use the coefficient of speed of sound $c_s^2$ to simulate different medium evolution. It is well known that when $c_s^2=1/3$ we get the evolution from one-dimensional ideal hydrodynamics, while to simulate the possibility to include viscous effects we employ a Full Relativistic Boltzmann Transport approach which has been developed over more than a decade in Catania to study collective bulk and heavy quarks dynamics in heavy-ion collisions for different RHIC and LHC systems~\cite{Plumari:2019hzp,Nugara:2024net,Nugara:2023eku,Nugara:2025ueb,Sambataro:2022sns,Sambataro:2023tlv,Sambataro:2025obe}. In particular, in this approach the cross section of bulk interaction is fixed in order to get relativistic hydrodynamic expansion at fixed viscous-to-entropy $\eta/s$ ratio.

In our model, the time evolution of the environment is encoded in a temperature-dependent variation of the rate of dissipation and decoherence induced by the system-reservoir interaction $\gamma(T)$. 
Different microscopic models predict different a different behavior for $\gamma(T)$. In this first study we stick to the $\gamma(T)$ parametrization from the original approach for charmonium suppression~\cite{Grandchamp:2002wp}.

\begin{figure}[h!]
    \centering
    \includegraphics[width=0.85\linewidth]{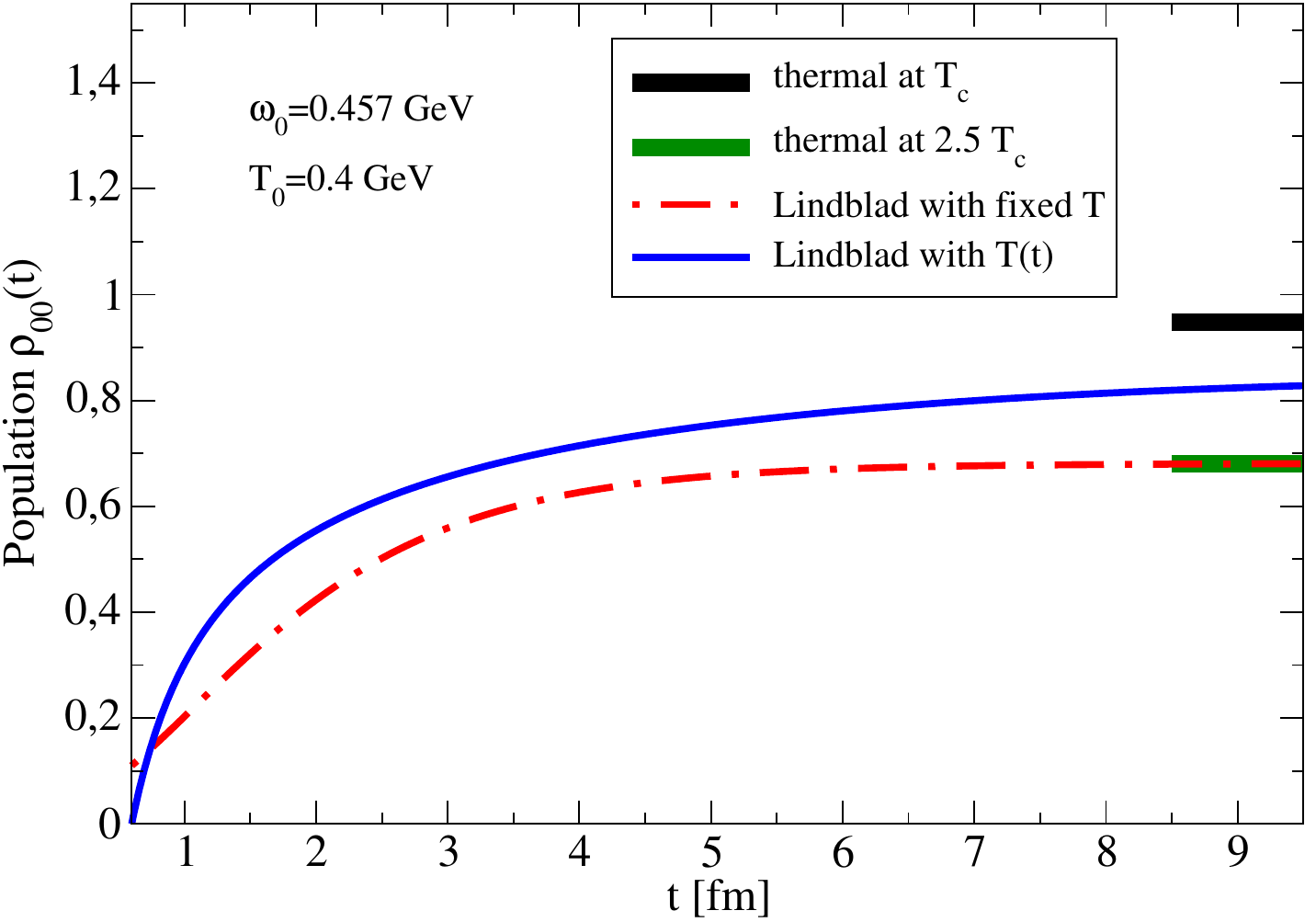}
    \caption{Time evolution of ground state population of $J/\Psi$ state ($\omega_0 = 0.457$ GeV) using Lindblad equation with static medium fixed temperature $T_0 = 0.4$ GeV and expanding medium with $T(t)$ according to Eq.~\eqref{eq:EvoTemp} with parameters set to 1D ideal hydrodynamics.}
    \label{fig:populationTevo1}
\end{figure}

In Fig.~\ref{fig:populationTevo1} we show the population of charmonium ($J/\Psi$) represented as the ground state of harmonic oscillator with frequency $\omega_0=0.457$ GeV as function of time obtained from the two different cases: static heat bath at temperature $T=2.5 T_c =0.4$ GeV (dot-dashed red line) and dynamical medium which expands from $T_0=0.4$ GeV and cools down according to one-dimensional hydrodynamical description. The two numerical results are confronted with the thermal values of $\rho_{00}$ at $T_c$ (black thick depth) and 2.5 $T_c$ (green thick depth). In this setup, we initialized the system with first excited state fully occupied $\rho_{11}(0)=1$.
As expected, for the $T(t)$-varying reservoir case the population of $J/\Psi$ ground state remains higher than the static one at fixed initial $T$, due to the fact that the bounded system, characterized by frequency $\omega_0$ increases as the temperature decreases. 
Of course, this difference depends strongly on the way the dissociation coefficient $\gamma(T)$ varies with temperature, as well as on our assumption that the frequency $\omega_0$ remains unaffected by the environment. 

\begin{figure}[h!]
    \centering
    \includegraphics[width=0.85\linewidth]{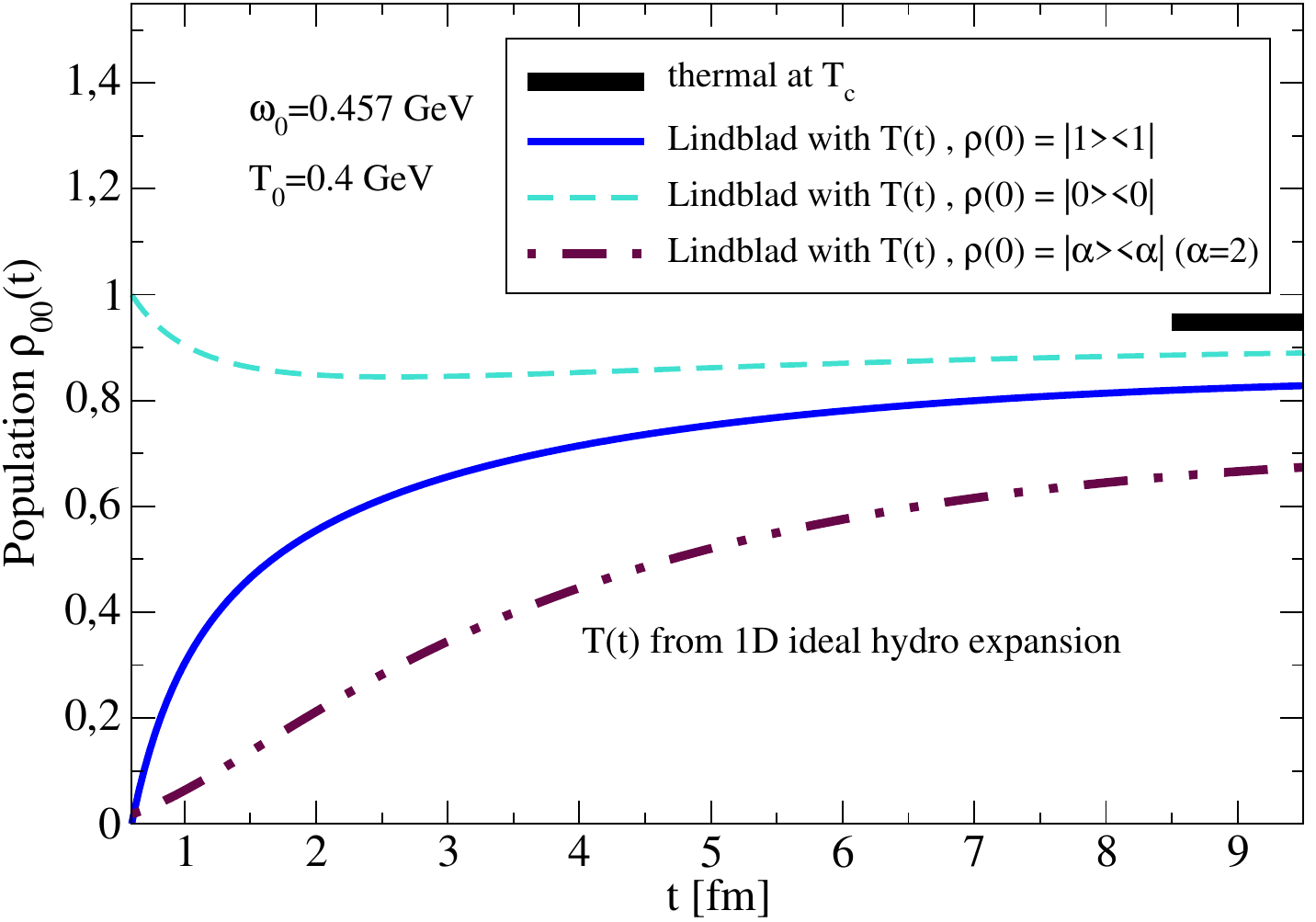}
    \caption{Time evolution of ground state population of $J/\Psi$ state ($\omega_0 = 0.457$ GeV) using Lindblad equation with $T(t)$ according to Eq.~\eqref{eq:EvoTemp} with parameters set to 1D ideal hydrodynamics. Various lines correspond to different initial conditions for the density matrix $\rho(0)$ and are compared with the thermal value (black thick line) as discussed in the text.}
    \label{fig:populationTevo1initcond}
\end{figure}

While in terms of population the situation of $T(t)$ environment seems quite similar to the static case, we observe in Fig.~\ref{fig:populationTevo1initcond} that considering the medium evolution through time-dependent coupling strength in the interaction Hamiltonian $H_{SR}(t)$ produces a slight dependence of the final population number on the system initial conditions.
Said so, we move to the study of quantum decoherence within the $T(t)$ case in comparison with the static heat bath by analyzing the time evolution of the off-diagonal terms of the reduced density matrix.
\begin{figure}[h!]
    \centering
    \includegraphics[width=\linewidth]{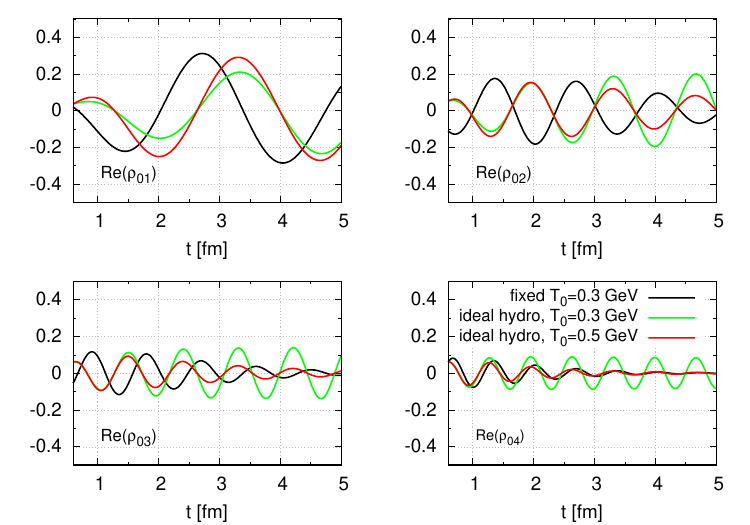}
    \caption{(color online): Time evolution of the real part of coherence elements $\rho_{n0}$ with $n=1,2,3,4$. Different colors correspond to different temperature for the heat bath: static medium at fixed $T=0.3$ GeV (black line), medium with temperature evolution $T(t)$ following ideal hydrodynamics at two different initial temperatures $T_0=0.3$ GeV (green line) and $T_0=0.5$ GeV (red line).}
    \label{fig:REcoherenceTevo1}
\end{figure}

\begin{figure}[h!]
    \centering
    \includegraphics[width=\linewidth]{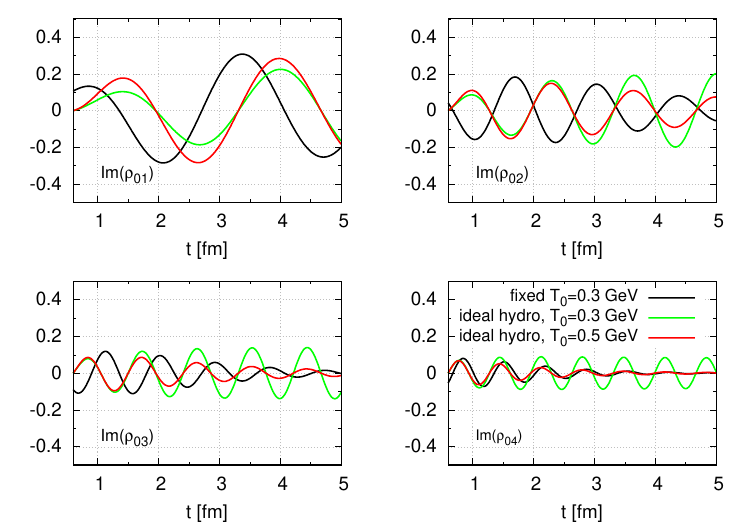}
    \caption{(color online): Time evolution of the imaginary part of coherence elements $\rho_{n0}$ with $n=1,2,3,4$. The description of the colored lines is the same of the plot for the real part Fig.~\ref{fig:REcoherenceTevo1}.}
    \label{fig:IMAGcoherenceTevo1}
\end{figure}

In Fig.~\ref{fig:REcoherenceTevo1} and~\ref{fig:IMAGcoherenceTevo1} we show respectively the real and imaginary part of four off-diagonal elements $\rho_{0n}$ with $n=1,\dots,4$ (from upper left to lower right) as function of time.
As initial condition of the system density matrix, we choose that of a pure coherent state $|\alpha \rangle \langle\alpha|$ of a single harmonic oscillator with characteristic frequency $\omega_0=0.457$ GeV (J/$\Psi$). The parameter $\alpha$ is set to 2 and the initialization as coherent state allows us to study the damped oscillations of coherences quite well.
As usual, the behavior of the imaginary part is shown for completeness but it follows the trend of the real part shifted by some constant phase. Focusing on Fig.~\ref{fig:REcoherenceTevo1} we compare the case of Lindblad evolution within a static heat bath at fixed $T=0.3$ GeV already presented in Fig.~\ref{fig:ReCoherenceT300} (black line) with two cases where we consider an evolving environment characterized by $T(t)$ dependence obtained from Eq.~\eqref{eq:EvoTemp} with $c_s^2=1/3$ (ideal hydrodynamics) for two different initial temperatures $T_0=0.3$ GeV (green line) and $T_0=0.5$ GeV (red line).

\begin{figure*}[t!]
    \includegraphics[width=\linewidth]{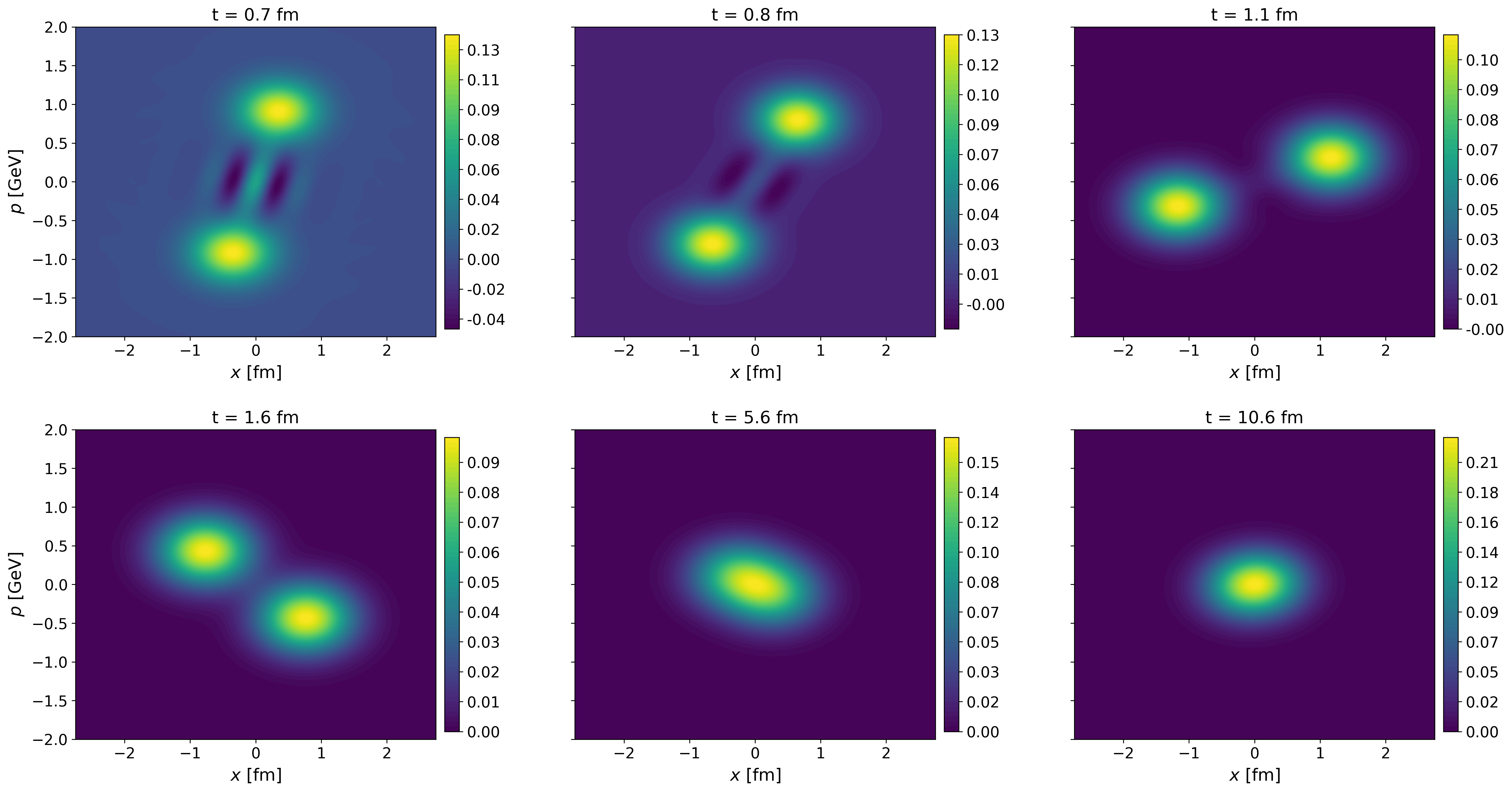}
    \caption{Contour plots of Wigner function $W(x,p,t)$ starting from initial cat state Eq.~\eqref{eq:catstate} with $\alpha=2$ and evolved according to the Lindblad equation Eq.~\eqref{eq:lindbGHJKL} in expanding medium using temperature $T(t)$ from Eq.~\eqref{eq:EvoTemp} with $T_0=0.5$ GeV and $c_s^2=1/3$ (1D ideal hydrodynamics).}
    \label{fig:wignerTtimeCat}
\end{figure*}

We observe that the damped oscillations in the off-diagonal elements exhibit a similar behavior when a time-dependent interaction strength between the system and the environment is considered. However, in the case of hydrodynamic expansion starting from $T_0 = 0.3$ GeV (green curve), the quantum coherences appear to remain more stable than in the case of a thermal bath with the same fixed initial temperature (black curve).
Therefore, if the initial temperature of the fireball is approximately that reached in RHIC collisions, the quantum decoherence process of quarkonia in the expanding medium could persist for times $t \gtrsim 5$ fm and probe the QGP evolution throughout the dissociation and recombination process of quarkonia.

\begin{figure}[h!]
    \centering
    \includegraphics[width=\linewidth]{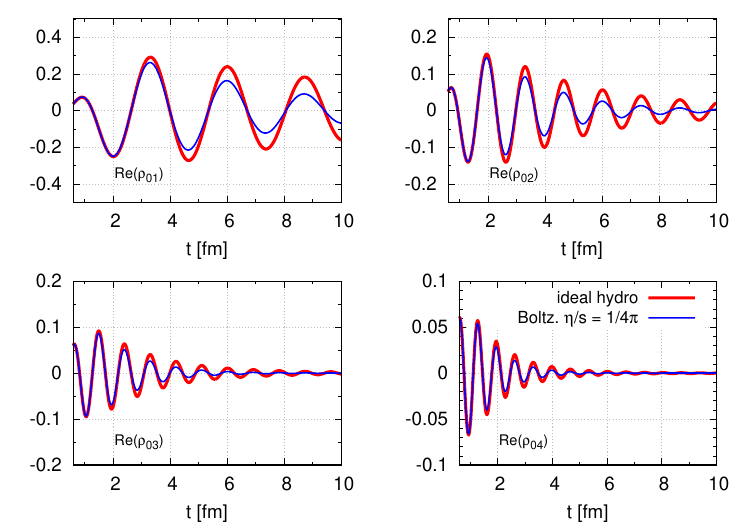}
    \caption{(color online) The real part of coherence elements $\rho_{n0}$ with $n=1,2,3,4$ (from upper left plot to lower right) for $J/\Psi$ system's Lindblad evolution in expanding QGP medium for two different scenario: $T(t)$ from 1D ideal hydro (thick red curve) and 1D kinetic Boltzmann at fixed $\eta/s=1/4\pi$ (thin blue curve).}
    \label{fig:CoherenceHydrovsBoltz}
\end{figure}

However, if the system starts from a higher initial temperature $T_0 = 0.5$ GeV (red curve), hence closer to the conditions achieved at the collisions of LHC, the decoherence process takes place on a much shorter timescale, $t \lesssim 2$ fm, effectively causing the system to acquire classical behavior.

Interestingly, we obtain a similar result for the case where the $Q\bar{Q}$ pair is initialized with a strongly correlated state such as the cat state Eq.~\eqref{eq:catstate} with parameter $\alpha=2$. In Fig.~\ref{fig:wignerTtimeCat} we show the profile of Wigner function $W(x,p,t)$ on the $xp$-plane at different times obtained from the Lindblad evolution of $\rho(t)$ within an expanding QGP medium where the temperature $T(t)$ varies according to Eq.~\eqref{eq:EvoTemp} for 1D ideal hydrodynamical fluid with $\tau_0=0.6$ fm, $T_0=0.5$ GeV and $c_s^2=1/3$. We observe that at about $t \simeq 1$ fm the fast decoherence process has removed all the negative fringes which were present in the initial Wigner function. Therefore, although $W(x,p,t)$ still shows two well-separated peaks and, hence, is very different from asymptotic gaussian-like distribution of Eq.~\eqref{eq:asymtoticWigner}, the vanishing of all initial quantum coherences allows the system to be considered at classical level. In this sense the Wigner function could be interpreted as the wavefunction of the hadronic state and convoluted with the heavy (anti-)quark single particle distribution as for example done in a coalescence hadronization approach which has been shown to provide the correct description of heavy-flavor spectra in $pp$ collisions~\cite{Minissale:2020bif,Minissale:2024gxx,Plumari:2025ptz}. We notice also that at $t \simeq 5$ fm, which is the typical lifetime of the fireball at RHIC and LHC energies, the shape of the Wigner function has become very close to the gaussian profile which is commonly employed in the literature to study heavy-flavor production by coalescence~\cite{Zhao:2023ucp,Zhao:2023nrz,Plumari:2017ntm,Minissale:2024gxx}.

In conclusion, we also investigate a scenario that includes viscous effects in the expansion of the QGP by using the time dependence of the temperature $T(t)$ obtained from the fully relativistic Boltzmann transport code from Catania~\cite{Nugara:2024net} which is tuned to reproduce a one-dimensional viscous hydrodynamic expansion with a fixed  $\eta/s = 1/4\pi$.
In Fig.~\ref{fig:CoherenceHydrovsBoltz}, we show the time evolution of the off-diagonal elements of the $J/\psi$ density matrix. The thick red curve corresponds to the hydrodynamic expansion case discussed previously, while the thin blue curve represents the 1D Boltzmann case with $\eta/s = 1/4\pi$. Both simulations try to describe the evolution of a QGP created at LHC conditions with $T_0 = 0.5$ GeV.
As can be seen, the two curves nearly overlap, suggesting that viscous effects do not significantly affect the decoherence of quarkonia.

\section{Conclusions}
In this study we have discussed the evolution of quarkonium-like states in heavy-ion collisions within an open quantum systems approach whose relevant effects are dissipation and decoherence. The density matrix of the system evolves according to quantum Lindblad master equation, which projected to phase-space representation is equivalent to a Fokker-Planck equation for the Wigner distribution.

In our work we have focused on quantum decoherence in the dynamics involving transitions between quarkonium states in strongly interacting matter. In line with other studies~\cite{Kajimoto:2017rel} we have shown that decoherence induced by quarkonium interaction with the QGP plays an important role as it drives the system towards a fast loss of quantum correlations, leading to the spontaneous emergence of the classical behavior which is commonly assumed in dynamical and hadronization models.

Moreover, we have extended the standard description based on the Lindblad master equation for a harmonic oscillator coupled to an environment through an amplitude-coupling interaction by developing a numerical framework in which the temperature of the reservoir evolves in time according to an ideal one-dimensional hydrodynamic expansion. We have found that the decoherence process is slowed down compared to the case of a static thermal bath. However, this result depends strongly on the initial conditions of the medium produced in heavy-ion collisions, while it seems that it is not sensible to viscosity effects.
This study represents a first step toward the development of a more realistic description of quarkonium dynamics in an evolving QGP.

In future, we plan to reformulate the same OQS approach changing the properties of the system into a more physical description of tightly-bound quarkonia states and to investigate different types of interactions which have been derived in lQCD or other theoretical and effective models.
To develop a full three-dimensional dynamical framework where decoherence of quarkonium is coupled also to coalescence model based Wigner density to study hadronization of heavy quark and anti-quark pairs and to obtain results which can be compared to experimental measurements will be part also of a future study.

\section{Acknowledgments}
 This work was financially supported by the PNRR MUR Project No. PE0000023-NQSTI. 
 The authors acknowledge Marco Ruggieri, Vincenzo Nugara, Vincenzo Minissale and Vincenzo Greco for the enriching discussions that improved this work.

\bibliography{biblio-quarkonium-oqs.bib}

@article{Nugara:2023eku,
    author = "Nugara, Vincenzo and Plumari, Salvatore and Oliva, Lucia and Greco, Vincenzo",
    title = "{Far-from-equilibrium attractors with full relativistic Boltzmann approach in boost-invariant and non-boost-invariant systems}",
    eprint = "2311.11921",
    archivePrefix = "arXiv",
    primaryClass = "hep-ph",
    doi = "10.1140/epjc/s10052-024-13227-1",
    journal = "Eur. Phys. J. C",
    volume = "84",
    number = "8",
    pages = "861",
    year = "2024"
}

@article{Nugara:2025ueb,
    author = "Nugara, Vincenzo and Borghini, Nicolas and Greco, Vincenzo and Plumari, Salvatore",
    title = "{Knudsen number and universal behavior of collective flows in conformal and non-conformal systems}",
    eprint = "2509.05495",
    archivePrefix = "arXiv",
    primaryClass = "hep-ph",
    doi = "10.1016/j.physletb.2025.140122",
    journal = "Phys. Lett. B",
    volume = "872",
    pages = "140122",
    year = "2026"
}

@article{Sambataro:2022sns,
    author = "Sambataro, Maria Lucia and Sun, Yifeng and Minissale, Vincenzo and Plumari, Salvatore and Greco, Vincenzo",
    title = "{Event-shape engineering analysis of D meson in ultrarelativistic heavy-ion collisions}",
    eprint = "2206.03160",
    archivePrefix = "arXiv",
    primaryClass = "hep-ph",
    doi = "10.1140/epjc/s10052-022-10802-2",
    journal = "Eur. Phys. J. C",
    volume = "82",
    number = "9",
    pages = "833",
    year = "2022"
}

@article{Sambataro:2023tlv,
    author = "Sambataro, Maria Lucia and Minissale, Vincenzo and Plumari, Salvatore and Greco, Vincenzo",
    title = "{B meson production in Pb+Pb at 5.02 ATeV at LHC: Estimating the diffusion coefficient in the infinite mass limit}",
    eprint = "2304.02953",
    archivePrefix = "arXiv",
    primaryClass = "hep-ph",
    doi = "10.1016/j.physletb.2024.138480",
    journal = "Phys. Lett. B",
    volume = "849",
    pages = "138480",
    year = "2024"
}

@article{Sambataro:2025obe,
    author = "Sambataro, Maria Lucia and Minissale, Vincenzo and Plumari, Salvatore and Greco, Vincenzo",
    title = "{Assessing lattice QCD charm space diffusion coefficient and thermalization time by mean of D meson observables at LHC}",
    eprint = "2508.01024",
    archivePrefix = "arXiv",
    primaryClass = "hep-ph",
    doi = "10.1016/j.physletb.2025.140049",
    journal = "Phys. Lett. B",
    volume = "872",
    pages = "140049",
    year = "2026"
}

@book{BreuerPetruccione2002,
  author    = {Heinz-Peter Breuer and Francesco Petruccione},
  title     = {The Theory of Open Quantum Systems},
  year      = {2002},
  publisher = {Oxford University Press},
  address   = {Oxford},
  isbn      = {978-0-19-921390-0}
}

@book{carmichael1993quantum,
  title     = {An Open Systems Approach to Quantum Optics},
  author    = {Carmichael, Howard J.},
  year      = {1993},
  publisher = {Springer},
  series    = {Lecture Notes in Physics},
  volume    = {m18},
  address   = {Berlin, Heidelberg},
  isbn      = {978-3-540-56689-1},
}

@book{weiss2008quantum,
  title     = {Quantum Dissipative Systems},
  author    = {Weiss, Ulrich},
  year      = {2008},
  publisher = {World Scientific},
  address   = {Singapore},
  isbn      = {978-9-812-79179-5},
}

@book{ScullyZubairy1997,
    author="Scully, Marlan O. and Zubairy, M. Suhail",
    title="{Quantum Optics}",
    publisher="Cambridge University Press", 
    year="1997"
}

@book{risken1989fpe,
  title={The Fokker-Planck Equation: Methods of Solution and Applications},
  author={Risken, Hannes},
  series={Springer Series in Synergetics},
  volume={18},
  edition={2nd},
  year={1989},
  publisher={Springer-Verlag},
  address={Berlin, Heidelberg},
  doi={10.1007/978-3-642-61544-3}
}

@article{Brasil_2013,
   title={A simple derivation of the Lindblad equation},
   volume={35},
   ISSN={1806-1117},
   url={http://dx.doi.org/10.1590/S1806-11172013000100003},
   DOI={10.1590/s1806-11172013000100003},
   number={1},
   journal={Revista Brasileira de Ensino de Física},
   publisher={FapUNIFESP (SciELO)},
   author={Brasil, Carlos Alexandre and Fanchini, Felipe Fernandes and Napolitano, Reginaldo de Jesus},
   year={2013},
   month=mar, pages={01–09} }

@article{Vidiella-Barranco:2016hnh,
    author = "Vidiella-Barranco, A.",
    title = "{Evolution of a quantum harmonic oscillator coupled to a minimal thermal environment}",
    eprint = "1605.01050",
    archivePrefix = "arXiv",
    primaryClass = "quant-ph",
    doi = "10.1016/j.physa.2016.04.033",
    journal = "Physica A",
    volume = "459",
    pages = "78",
    year = "2016"
}

@article{Coci:2025drb,
    author = "Coci, Gabriele and Parisi, Gabriele and Plumari, Salvatore and Ruggieri, Marco",
    title = "{Entropy from decoherence: a case study using glasma-based occupation numbers}",
    eprint = "2507.04809",
    archivePrefix = "arXiv",
    primaryClass = "hep-ph",
    month = "7",
    year = "2025"
}

@article{Lindblad:1975ef,
    author = "Lindblad, Goran",
    title = "{On the Generators of Quantum Dynamical Semigroups}",
    reportNumber = "TRITA-TFY-75-1",
    doi = "10.1007/BF01608499",
    journal = "Commun. Math. Phys.",
    volume = "48",
    pages = "119",
    year = "1976"
}

@article{Gorini:1975nb,
    author = "Gorini, Vittorio and Kossakowski, Andrzej and Sudarshan, E. C. G.",
    title = "{Completely Positive Dynamical Semigroups of N Level Systems}",
    reportNumber = "CPT-244-TEXAS, ORO-3992-200",
    doi = "10.1063/1.522979",
    journal = "J. Math. Phys.",
    volume = "17",
    pages = "821",
    year = "1976"
}

@article{Rais:2025fps,
    author = "Rais, Jan and van Hees, Hendrik and Greiner, Carsten",
    title = "{Bound-state formation and thermalization within the Lindblad approach}",
    doi = "10.1103/PhysRevC.111.054918",
    journal = "Phys. Rev. C",
    volume = "111",
    number = "5",
    pages = "054918",
    year = "2025"
}

@article{Neidig:2023kid,
    author = "Neidig, Tim and Rais, Jan and Bleicher, Marcus and van Hees, Hendrik and Greiner, Carsten",
    title = "{Open quantum systems with Kadanoff-Baym equations}",
    eprint = "2308.07659",
    archivePrefix = "arXiv",
    primaryClass = "nucl-th",
    doi = "10.1016/j.physletb.2024.138589",
    journal = "Phys. Lett. B",
    volume = "851",
    pages = "138589",
    year = "2024"
}

@article{caldeira1983quantum,
  title={Quantum tunnelling in a dissipative system},
  author={Caldeira, Amir O and Leggett, Anthony J},
  journal={Annals of physics},
  volume={149},
  number={2},
  pages={374--456},
  year={1983},
  publisher={Academic Press}
}

@article{Caldeira:1982iu,
    author = "Caldeira, A. O. and Leggett, A. J.",
    title = "{Path integral approach to quantum Brownian motion}",
    doi = "10.1016/0378-4371(83)90013-4",
    journal = "Physica A",
    volume = "121",
    pages = "587--616",
    year = "1983"
}

@article{Schlosshauer:2019ewh,
    author = "Schlosshauer, Maximilian",
    title = "{Quantum decoherence}",
    eprint = "1911.06282",
    archivePrefix = "arXiv",
    primaryClass = "quant-ph",
    doi = "10.1016/j.physrep.2019.10.001",
    journal = "Phys. Rept.",
    volume = "831",
    pages = "1--57",
    year = "2019"
}

@article{Schlosshauer:2003zy,
    author = "Schlosshauer, Maximilian",
    title = "{Decoherence, the Measurement Problem, and Interpretations of Quantum Mechanics}",
    eprint = "quant-ph/0312059",
    archivePrefix = "arXiv",
    doi = "10.1103/RevModPhys.76.1267",
    journal = "Rev. Mod. Phys.",
    volume = "76",
    pages = "1267--1305",
    year = "2004"
}

@article{Zurek:2003zz,
    author = "Zurek, Wojciech Hubert",
    title = "{Decoherence, einselection, and the quantum origins of the classical}",
    eprint = "quant-ph/0105127",
    archivePrefix = "arXiv",
    doi = "10.1103/RevModPhys.75.715",
    journal = "Rev. Mod. Phys.",
    volume = "75",
    pages = "715--775",
    year = "2003"
}

@article{Zurek:1991vd,
    author = "Zurek, Wojciech H.",
    title = "{Decoherence and the transition from quantum to classical}",
    eprint = "quant-ph/0306072",
    archivePrefix = "arXiv",
    doi = "10.1063/1.881293",
    journal = "Phys. Today",
    volume = "44N10",
    pages = "36--44",
    year = "1991"
}

@article{Song:2024got,
    author = "Song, Taesoo and Aichelin, Joerg and Zhao, Jiaxing and Gossiaux, Pol B. and Bratkovskaya, Elena",
    title = "{Quarkonium production in pp and heavy-ion collisions}",
    eprint = "2409.19280",
    archivePrefix = "arXiv",
    primaryClass = "hep-ph",
    doi = "10.1051/epjconf/202531604008",
    journal = "EPJ Web Conf.",
    volume = "316",
    pages = "04008",
    year = "2025"
}

@article{Yao:2025jyx,
    author = "Yao, Xiaojun",
    title = "{Quarkonia Theory: From Open Quantum System to Classical Transport}",
    eprint = "2505.00098",
    archivePrefix = "arXiv",
    primaryClass = "hep-ph",
    reportNumber = "IQuS@UW-21-101",
    doi = "10.1051/epjconf/202533901013",
    journal = "EPJ Web Conf.",
    volume = "339",
    pages = "01013",
    year = "2025"
}

@article{Yao:2020xzw,
    author = {Yao, Xiaojun and Ke, Weiyao and Xu, Yingru and Bass, Steffen A. and M{\"u}ller, Berndt},
    title = "{Coupled Boltzmann Transport Equations of Heavy Quarks and Quarkonia in Quark-Gluon Plasma}",
    eprint = "2004.06746",
    archivePrefix = "arXiv",
    primaryClass = "hep-ph",
    reportNumber = "MIT-CTP/5192",
    doi = "10.1007/JHEP01(2021)046",
    journal = "JHEP",
    volume = "01",
    pages = "046",
    year = "2021"
}

@article{Zhao:2020jqu,
    author = "Zhao, Jiaxing and Zhou, Kai and Chen, Shile and Zhuang, Pengfei",
    title = "{Heavy flavors under extreme conditions in high energy nuclear collisions}",
    eprint = "2005.08277",
    archivePrefix = "arXiv",
    primaryClass = "nucl-th",
    doi = "10.1016/j.ppnp.2020.103801",
    journal = "Prog. Part. Nucl. Phys.",
    volume = "114",
    pages = "103801",
    year = "2020"
}

@inproceedings{Rapp:2009my,
    author = "Rapp, Ralf and van Hees, Hendrik",
    title = "{Heavy Quarks in the Quark-Gluon Plasma}",
    eprint = "0903.1096",
    archivePrefix = "arXiv",
    primaryClass = "hep-ph",
    doi = "10.1142/9789814293297_0003",
    pages = "111--206",
    year = "2010"
}

@article{Dong:2019unq,
    author = "Dong, Xin and Greco, Vincenzo",
    title = "{Heavy quark production and properties of Quark{\textendash}Gluon Plasma}",
    doi = "10.1016/j.ppnp.2018.08.001",
    journal = "Prog. Part. Nucl. Phys.",
    volume = "104",
    pages = "97--141",
    year = "2019"
}

@article{Rothkopf:2019ipj,
    author = "Rothkopf, Alexander",
    title = "{Heavy Quarkonium in Extreme Conditions}",
    eprint = "1912.02253",
    archivePrefix = "arXiv",
    primaryClass = "hep-ph",
    doi = "10.1016/j.physrep.2020.02.006",
    journal = "Phys. Rept.",
    volume = "858",
    pages = "1--117",
    year = "2020"
}

@article{Andronic:2015wma,
    author = "Andronic, A. and others",
    title = "{Heavy-flavour and quarkonium production in the LHC era: from proton{\textendash}proton to heavy-ion collisions}",
    eprint = "1506.03981",
    archivePrefix = "arXiv",
    primaryClass = "nucl-ex",
    doi = "10.1140/epjc/s10052-015-3819-5",
    journal = "Eur. Phys. J. C",
    volume = "76",
    number = "3",
    pages = "107",
    year = "2016"
}

@article{NA50:1996lag,
    author = "Gonin, Michel and others",
    editor = "Braun-Munzinger, P. and Specht, H. J. and Stock, R. and Stoecker, Horst",
    collaboration = "NA50",
    title = "{Anomalous J / psi suppression in Pb + Pb collisions at 158-A-GeV/c}",
    doi = "10.1016/S0375-9474(96)00373-9",
    journal = "Nucl. Phys. A",
    volume = "610",
    pages = "404C--417C",
    year = "1996"
}

@article{NA60:2006ncq,
    author = "Arnaldi, R. and others",
    editor = "Sissakian, Alexey and Kozlov, Gennady and Kolganova, Elena",
    collaboration = "NA60",
    title = "{$J/\psi$ production in Indium-Indium collisions at 158- GeV/nucleon}",
    eprint = "0706.4361",
    archivePrefix = "arXiv",
    primaryClass = "nucl-ex",
    doi = "10.1103/PhysRevLett.99.132302",
    journal = "Conf. Proc. C",
    volume = "060726",
    pages = "430--434",
    year = "2006"
}

@article{PHENIX:2006gsi,
    author = "Adare, A. and others",
    collaboration = "PHENIX",
    title = "{$J/\psi$ Production vs Centrality, Transverse Momentum, and Rapidity in Au+Au Collisions at $\sqrt{s_{NN}} = 200$ GeV}",
    eprint = "nucl-ex/0611020",
    archivePrefix = "arXiv",
    doi = "10.1103/PhysRevLett.98.232301",
    journal = "Phys. Rev. Lett.",
    volume = "98",
    pages = "232301",
    year = "2007"
}

@article{STAR:2009irl,
    author = "Abelev, B. I. and others",
    collaboration = "STAR",
    title = "{J/psi production at high transverse momentum in p+p and Cu+Cu collisions at s(NN)**1/2 = 200GeV}",
    eprint = "0904.0439",
    archivePrefix = "arXiv",
    primaryClass = "nucl-ex",
    doi = "10.1103/PhysRevC.80.041902",
    journal = "Phys. Rev. C",
    volume = "80",
    pages = "041902",
    year = "2009"
}

@article{STAR:2013kwk,
    author = "Adamczyk, L. and others",
    collaboration = "STAR",
    title = "{Suppression of $\Upsilon$ production in d+Au and Au+Au collisions at $\sqrt{s_{NN}}$=200 GeV}",
    eprint = "1312.3675",
    archivePrefix = "arXiv",
    primaryClass = "nucl-ex",
    doi = "10.1016/j.physletb.2014.06.028",
    journal = "Phys. Lett. B",
    volume = "735",
    pages = "127--137",
    year = "2014",
    note = "[Erratum: Phys.Lett.B 743, 537--541 (2015)]"
}

@article{ALICE:2012jsl,
    author = "Abelev, Betty and others",
    collaboration = "ALICE",
    title = "{$J/\psi$ suppression at forward rapidity in Pb-Pb collisions at $\sqrt{s_{NN}}=2.76$ TeV}",
    eprint = "1202.1383",
    archivePrefix = "arXiv",
    primaryClass = "hep-ex",
    reportNumber = "CERN-PH-EP-2012-012",
    doi = "10.1103/PhysRevLett.109.072301",
    journal = "Phys. Rev. Lett.",
    volume = "109",
    pages = "072301",
    year = "2012"
}

@article{CMS:2012gvv,
    author = "Chatrchyan, Serguei and others",
    collaboration = "CMS",
    title = "{Observation of Sequential Upsilon Suppression in PbPb Collisions}",
    eprint = "1208.2826",
    archivePrefix = "arXiv",
    primaryClass = "nucl-ex",
    reportNumber = "CMS-HIN-11-011, CERN-PH-EP-2012-228",
    doi = "10.1103/PhysRevLett.109.222301",
    journal = "Phys. Rev. Lett.",
    volume = "109",
    pages = "222301",
    year = "2012",
    note = "[Erratum: Phys.Rev.Lett. 120, 199903 (2018)]"
}

@article{ALICE:2018bdo,
    author = "Acharya, Shreyasi and others",
    collaboration = "ALICE",
    title = "{Study of J/$\psi$ azimuthal anisotropy at forward rapidity in Pb-Pb collisions at $ \sqrt{s_{\mathrm{NN}}}=5.02 $ TeV}",
    eprint = "1811.12727",
    archivePrefix = "arXiv",
    primaryClass = "nucl-ex",
    reportNumber = "CERN-EP-2018-319",
    doi = "10.1007/JHEP02(2019)012",
    journal = "JHEP",
    volume = "02",
    pages = "012",
    year = "2019"
}

@article{ParticleDataGroup:2024cfk,
    author = "Navas, S. and others",
    collaboration = "Particle Data Group",
    title = "{Review of particle physics}",
    doi = "10.1103/PhysRevD.110.030001",
    journal = "Phys. Rev. D",
    volume = "110",
    number = "3",
    pages = "030001",
    year = "2024"
}

@article{Matsui:1986dk,
    author = "Matsui, T. and Satz, H.",
    title = "{$J/\psi$ Suppression by Quark-Gluon Plasma Formation}",
    reportNumber = "BNL-38344",
    doi = "10.1016/0370-2693(86)91404-8",
    journal = "Phys. Lett. B",
    volume = "178",
    pages = "416--422",
    year = "1986"
}

@article{Andronic:2024oxz,
    author = "Andronic, A. and others",
    title = "{Comparative study of quarkonium transport in hot QCD matter}",
    eprint = "2402.04366",
    archivePrefix = "arXiv",
    primaryClass = "nucl-th",
    reportNumber = "FERMILAB-PUB-24-0005-T-V",
    doi = "10.1140/epja/s10050-024-01306-6",
    journal = "Eur. Phys. J. A",
    volume = "60",
    number = "4",
    pages = "88",
    year = "2024"
}

@article{Brambilla:1999xf,
    author = "Brambilla, Nora and Pineda, Antonio and Soto, Joan and Vairo, Antonio",
    title = "{Potential NRQCD: An Effective theory for heavy quarkonium}",
    eprint = "hep-ph/9907240",
    archivePrefix = "arXiv",
    reportNumber = "CERN-TH-99-199, HEPHY-PUB-716-99, UB-ECM-PF-99-06, UWTHPH-1999-34, UB-ECM-PF-99-13",
    doi = "10.1016/S0550-3213(99)00693-8",
    journal = "Nucl. Phys. B",
    volume = "566",
    pages = "275",
    year = "2000"
}

@article{Petreczky:2008px,
    author = "Petreczky, P.",
    title = "{On temperature dependence of quarkonium correlators}",
    eprint = "0810.0258",
    archivePrefix = "arXiv",
    primaryClass = "hep-lat",
    doi = "10.1140/epjc/s10052-009-0942-1",
    journal = "Eur. Phys. J. C",
    volume = "62",
    pages = "85--93",
    year = "2009"
}

@article{Brambilla:2022ynh,
    author = "Brambilla, Nora and Escobedo, Miguel \'Angel and Islam, Ajaharul and Strickland, Michael and Tiwari, Anurag and Vairo, Antonio and Vander Griend, Peter",
    title = "{Heavy quarkonium dynamics at next-to-leading order in the binding energy over temperature}",
    eprint = "2205.10289",
    archivePrefix = "arXiv",
    primaryClass = "hep-ph",
    reportNumber = "TUM-EFT 169/22",
    doi = "10.1007/JHEP08(2022)303",
    journal = "JHEP",
    volume = "08",
    pages = "303",
    year = "2022"
}

@article{Brambilla:2020qwo,
    author = "Brambilla, Nora and Escobedo, Miguel \'Angel and Strickland, Michael and Vairo, Antonio and Vander Griend, Peter and Weber, Johannes Heinrich",
    title = "{Bottomonium suppression in an open quantum system using the quantum trajectories method}",
    eprint = "2012.01240",
    archivePrefix = "arXiv",
    primaryClass = "hep-ph",
    reportNumber = "TUM-EFT 140/20; HU-EP-20/36-RTG",
    doi = "10.1007/JHEP05(2021)136",
    journal = "JHEP",
    volume = "05",
    pages = "136",
    year = "2021"
}

@article{Brambilla:2017zei,
    author = "Brambilla, Nora and Escobedo, Miguel A. and Soto, Joan and Vairo, Antonio",
    title = "{Heavy quarkonium suppression in a fireball}",
    eprint = "1711.04515",
    archivePrefix = "arXiv",
    primaryClass = "hep-ph",
    reportNumber = "TUM-EFT-89-16",
    doi = "10.1103/PhysRevD.97.074009",
    journal = "Phys. Rev. D",
    volume = "97",
    number = "7",
    pages = "074009",
    year = "2018"
}

@article{Brambilla:2016wgg,
    author = "Brambilla, Nora and Escobedo, Miguel A. and Soto, Joan and Vairo, Antonio",
    title = "{Quarkonium suppression in heavy-ion collisions: an open quantum system approach}",
    eprint = "1612.07248",
    archivePrefix = "arXiv",
    primaryClass = "hep-ph",
    reportNumber = "ICCUB-16-044, TUM-EFT-55-14",
    doi = "10.1103/PhysRevD.96.034021",
    journal = "Phys. Rev. D",
    volume = "96",
    number = "3",
    pages = "034021",
    year = "2017"
}

@article{Akamatsu:2011se,
    author = "Akamatsu, Yukinao and Rothkopf, Alexander",
    title = "{Stochastic potential and quantum decoherence of heavy quarkonium in the quark-gluon plasma}",
    eprint = "1110.1203",
    archivePrefix = "arXiv",
    primaryClass = "hep-ph",
    reportNumber = "BI-TP-2011-32",
    doi = "10.1103/PhysRevD.85.105011",
    journal = "Phys. Rev. D",
    volume = "85",
    pages = "105011",
    year = "2012"
}

@article{Akamatsu:2012vt,
    author = "Akamatsu, Yukinao",
    title = "{Real-time quantum dynamics of heavy quark systems at high temperature}",
    eprint = "1209.5068",
    archivePrefix = "arXiv",
    primaryClass = "hep-ph",
    doi = "10.1103/PhysRevD.87.045016",
    journal = "Phys. Rev. D",
    volume = "87",
    number = "4",
    pages = "045016",
    year = "2013"
}

@article{Akamatsu:2014qsa,
    author = "Akamatsu, Yukinao",
    title = "{Heavy quark master equations in the Lindblad form at high temperatures}",
    eprint = "1403.5783",
    archivePrefix = "arXiv",
    primaryClass = "hep-ph",
    doi = "10.1103/PhysRevD.91.056002",
    journal = "Phys. Rev. D",
    volume = "91",
    number = "5",
    pages = "056002",
    year = "2015"
}

@article{Kajimoto:2017rel,
    author = "Kajimoto, Shiori and Akamatsu, Yukinao and Asakawa, Masayuki and Rothkopf, Alexander",
    title = "{Dynamical dissociation of quarkonia by wave function decoherence}",
    eprint = "1705.03365",
    archivePrefix = "arXiv",
    primaryClass = "nucl-th",
    reportNumber = "PHYS.-REV.-D-97, 014003-(2018)",
    doi = "10.1103/PhysRevD.97.014003",
    journal = "Phys. Rev. D",
    volume = "97",
    number = "1",
    pages = "014003",
    year = "2018"
}

@article{Miura:2019ssi,
    author = "Miura, Takahiro and Akamatsu, Yukinao and Asakawa, Masayuki and Rothkopf, Alexander",
    title = "{Quantum Brownian motion of a heavy quark pair in the quark-gluon plasma}",
    eprint = "1908.06293",
    archivePrefix = "arXiv",
    primaryClass = "nucl-th",
    doi = "10.1103/PhysRevD.101.034011",
    journal = "Phys. Rev. D",
    volume = "101",
    number = "3",
    pages = "034011",
    year = "2020"
}

@article{Borghini:2011ms,
    author = "Borghini, Nicolas and Gombeaud, Clement",
    title = "{Heavy quarkonia in a medium as a quantum dissipative system: Master equation approach}",
    eprint = "1109.4271",
    archivePrefix = "arXiv",
    primaryClass = "nucl-th",
    reportNumber = "BI-TP-2011-031",
    doi = "10.1140/epjc/s10052-012-2000-7",
    journal = "Eur. Phys. J. C",
    volume = "72",
    pages = "2000",
    year = "2012"
}

@article{Katz:2015qja,
    author = "Katz, Roland and Gossiaux, Pol Bernard",
    title = {{The Schr\"odinger\textendash{}Langevin equation with and without thermal fluctuations}},
    eprint = "1504.08087",
    archivePrefix = "arXiv",
    primaryClass = "quant-ph",
    doi = "10.1016/j.aop.2016.02.005",
    journal = "Annals Phys.",
    volume = "368",
    pages = "267--295",
    year = "2016"
}

@article{Blaizot:2015hya,
    author = "Blaizot, Jean-Paul and De Boni, Davide and Faccioli, Pietro and Garberoglio, Giovanni",
    title = "{Heavy quark bound states in a quark\textendash{}gluon plasma: Dissociation and recombination}",
    eprint = "1503.03857",
    archivePrefix = "arXiv",
    primaryClass = "nucl-th",
    doi = "10.1016/j.nuclphysa.2015.10.011",
    journal = "Nucl. Phys. A",
    volume = "946",
    pages = "49--88",
    year = "2016"
}

@article{Blaizot:2017ypk,
    author = "Blaizot, Jean-Paul and Escobedo, Miguel Angel",
    title = "{Quantum and classical dynamics of heavy quarks in a quark-gluon plasma}",
    eprint = "1711.10812",
    archivePrefix = "arXiv",
    primaryClass = "hep-ph",
    doi = "10.1007/JHEP06(2018)034",
    journal = "JHEP",
    volume = "06",
    pages = "034",
    year = "2018"
}

@article{Delorme:2024rdo,
    author = "Delorme, St{\'e}phane and Katz, Roland and Gousset, Thierry and Gossiaux, Pol Bernard and Blaizot, Jean-Paul",
    title = "{Quarkonium dynamics in the quantum Brownian regime with non-abelian quantum master equations}",
    eprint = "2402.04488",
    archivePrefix = "arXiv",
    primaryClass = "hep-ph",
    reportNumber = "IFJPAN-IV-2024-3",
    doi = "10.1007/JHEP06(2024)060",
    journal = "JHEP",
    volume = "06",
    pages = "060",
    year = "2024"
}

@article{DeJong:2020riy,
    author = "De Jong, Wibe A. and Metcalf, Mekena and Mulligan, James and P\l{}osko\'n, Mateusz and Ringer, Felix and Yao, Xiaojun",
    title = "{Quantum simulation of open quantum systems in heavy-ion collisions}",
    eprint = "2010.03571",
    archivePrefix = "arXiv",
    primaryClass = "hep-ph",
    reportNumber = "MIT-CTP/5247",
    doi = "10.1103/PhysRevD.104.L051501",
    journal = "Phys. Rev. D",
    volume = "104",
    number = "5",
    pages = "051501",
    year = "2021"
}

@article{Grandchamp:2002wp,
    author = "Grandchamp, L. and Rapp, R.",
    title = "{Charmonium suppression and regeneration from SPS to RHIC}",
    eprint = "hep-ph/0205305",
    archivePrefix = "arXiv",
    doi = "10.1016/S0375-9474(02)01027-8",
    journal = "Nucl. Phys. A",
    volume = "709",
    pages = "415--439",
    year = "2002"
}

@article{Grandchamp:2001pf,
    author = "Grandchamp, L. and Rapp, R.",
    title = "{Thermal versus direct J / Psi production in ultrarelativistic heavy ion collisions}",
    eprint = "hep-ph/0103124",
    archivePrefix = "arXiv",
    reportNumber = "SUNY-NTG-01-4",
    doi = "10.1016/S0370-2693(01)01311-9",
    journal = "Phys. Lett. B",
    volume = "523",
    pages = "60--66",
    year = "2001"
}

@article{Song:2023zma,
    author = "Song, Taesoo and Aichelin, Joerg and Zhao, Jiaxing and Gossiaux, Pol Bernard and Bratkovskaya, Elena",
    title = "{Bottomonium production in pp and heavy-ion collisions}",
    eprint = "2305.10750",
    archivePrefix = "arXiv",
    primaryClass = "nucl-th",
    doi = "10.1103/PhysRevC.108.054908",
    journal = "Phys. Rev. C",
    volume = "108",
    number = "5",
    pages = "054908",
    year = "2023"
}

@article{Laine:2006ns,
    author = "Laine, M. and Philipsen, O. and Romatschke, P. and Tassler, M.",
    title = "{Real-time static potential in hot QCD}",
    eprint = "hep-ph/0611300",
    archivePrefix = "arXiv",
    reportNumber = "BI-TP-2006-41, MS-TP-06-32, INT-PUB-06-37",
    doi = "10.1088/1126-6708/2007/03/054",
    journal = "JHEP",
    volume = "03",
    pages = "054",
    year = "2007"
}

@article{Walls:1985tm,
    author = "Walls, D. F. and Milburn, G. J.",
    title = "{EFFECT OF DISSIPATION ON QUANTUM COHERENCE}",
    reportNumber = "NSF-ITP-85-65",
    month = "8",
    year = "1985"
}

@article{Yurke:1986zz,
    author = "Yurke, B. and Stoler, D.",
    title = "{Generating quantum mechanical superpositions of macroscopically distinguishable states via amplitude dispersion}",
    doi = "10.1103/PhysRevLett.57.13",
    journal = "Phys. Rev. Lett.",
    volume = "57",
    pages = "13--16",
    year = "1986"
}

@article{Saito:2003hv,
	author = "Saito, Y. and Hyuga, H.",
    title = "{Relaxation of Schroedinger cat states and displaced thermal states in a density operator representation}",
    journal = "Journal of the Physical Society of Japan",
    volume = "65",
    pages = "1648-1654",
    year = "1996"
}

@article{Cahill:1969it,
    author = "Cahill, Kevin E. and Glauber, R. J.",
    title = "{Ordered expansions in boson amplitude operators}",
    doi = "10.1103/PhysRev.177.1857",
    journal = "Phys. Rev.",
    volume = "177",
    pages = "1857--1881",
    year = "1969"
}

@article{Mehtar-Tani:2025xxd,
    author = "Mehtar-Tani, Yacine and Ringer, Felix and Singh, Balbeer and Vaidya, Varun",
    title = "{Open quantum system approach to inclusive jet production in heavy-ion collisions}",
    eprint = "2504.00101",
    archivePrefix = "arXiv",
    primaryClass = "hep-ph",
    doi = "10.1007/JHEP02(2026)048",
    journal = "JHEP",
    volume = "02",
    pages = "048",
    year = "2026"
}

@article{Iida:2014wea,
    author = "Iida, Hideaki and Kunihiro, Teiji and Ohnishi, Akira and Takahashi, Toru T.",
    title = "{Time evolution of gluon coherent state and its von Neumann entropy in heavy-ion collisions}",
    eprint = "1410.7309",
    archivePrefix = "arXiv",
    primaryClass = "hep-ph",
    reportNumber = "KUNS-2526, YITP-14-84",
    month = "10",
    year = "2014"
}

@article{Magnus:1954zz,
    author = "Magnus, Wilhelm",
    title = "{On the exponential solution of differential equations for a linear operator}",
    doi = "10.1002/cpa.3160070404",
    journal = "Commun. Pure Appl. Math.",
    volume = "7",
    pages = "649--673",
    year = "1954"
}

@article{Bjorken:1982qr,
    author = "Bjorken, J. D.",
    title = "{Highly Relativistic Nucleus-Nucleus Collisions: The Central Rapidity Region}",
    reportNumber = "FERMILAB-PUB-82-044-THY, FERMILAB-PUB-82-044-T",
    doi = "10.1103/PhysRevD.27.140",
    journal = "Phys. Rev. D",
    volume = "27",
    pages = "140--151",
    year = "1983"
}

@article{Plumari:2019hzp,
    author = "Plumari, Salvatore and Coci, Gabriele and Minissale, Vincenzo and Das, Santosh K. and Sun, Yifeng and Greco, Vincenzo",
    title = "{Heavy - light flavor correlations of anisotropic flows at LHC energies within event-by-event transport approach}",
    eprint = "1912.09350",
    archivePrefix = "arXiv",
    primaryClass = "hep-ph",
    doi = "10.1016/j.physletb.2020.135460",
    journal = "Phys. Lett. B",
    volume = "805",
    pages = "135460",
    year = "2020"
}

@article{Minissale:2024gxx,
    author = "Minissale, Vincenzo and Greco, Vincenzo and Plumari, Salvatore",
    title = "{Bottomed mesons and baryons production in pp collisions at s=5 TeV LHC energy within a Coalescence plus Fragmentation approach}",
    eprint = "2405.19244",
    archivePrefix = "arXiv",
    primaryClass = "hep-ph",
    doi = "10.1016/j.physletb.2024.139190",
    journal = "Phys. Lett. B",
    volume = "860",
    pages = "139190",
    year = "2025"
}

@article{Minissale:2023dct,
    author = "Minissale, Vincenzo and Plumari, Salvatore and Sun, Yifeng and Greco, Vincenzo",
    title = "{Multi-charmed and singled charmed hadrons from coalescence: yields and ratios in different collision systems at LHC}",
    eprint = "2305.03687",
    archivePrefix = "arXiv",
    primaryClass = "hep-ph",
    doi = "10.1140/epjc/s10052-024-12571-6",
    journal = "Eur. Phys. J. C",
    volume = "84",
    number = "3",
    pages = "228",
    year = "2024"
}

@article{Plumari:2017ntm,
    author = "Plumari, Salvatore and Minissale, Vincenzo and Das, Santosh K. and Coci, G. and Greco, V.",
    title = "{Charmed Hadrons from Coalescence plus Fragmentation in relativistic nucleus-nucleus collisions at RHIC and LHC}",
    eprint = "1712.00730",
    archivePrefix = "arXiv",
    primaryClass = "hep-ph",
    doi = "10.1140/epjc/s10052-018-5828-7",
    journal = "Eur. Phys. J. C",
    volume = "78",
    number = "4",
    pages = "348",
    year = "2018"
}

@article{Nugara:2024net,
    author = "Nugara, Vincenzo and Greco, Vincenzo and Plumari, Salvatore",
    title = "{Far-from-equilibrium attractors with Full Relativistic Boltzmann approach in 3+1D: moments of distribution function and~anisotropic flows $v_n$}",
    eprint = "2409.12123",
    archivePrefix = "arXiv",
    primaryClass = "hep-ph",
    doi = "10.1140/epjc/s10052-025-14029-9",
    journal = "Eur. Phys. J. C",
    volume = "85",
    number = "3",
    pages = "311",
    year = "2025"
}

@article{Minissale:2020bif,
    author = "Minissale, Vincenzo and Plumari, Salvatore and Greco, Vincenzo",
    title = "{Charm hadrons in pp collisions at LHC energy within a coalescence plus fragmentation approach}",
    eprint = "2012.12001",
    archivePrefix = "arXiv",
    primaryClass = "hep-ph",
    doi = "10.1016/j.physletb.2021.136622",
    journal = "Phys. Lett. B",
    volume = "821",
    pages = "136622",
    year = "2021"
}

@article{Plumari:2025ptz,
    author = "Plumari, Salvatore and Minissale, Vincenzo and Greco, Vincenzo",
    title = "{Charm and Bottom hadron production with a coalescence plus fragmentation hadronization approach: AA system size scan down to pp collisions}",
    doi = "10.1051/epjconf/202531604002",
    journal = "EPJ Web Conf.",
    volume = "316",
    pages = "04002",
    year = "2025"
}

@article{Zhao:2023ucp,
    author = "Zhao, Jiaxing and Aichelin, Joerg and Gossiaux, Pol Bernard and Werner, Klaus",
    title = "{Heavy flavor as a probe of hot QCD matter produced in proton-proton collisions}",
    eprint = "2310.08684",
    archivePrefix = "arXiv",
    primaryClass = "hep-ph",
    doi = "10.1103/PhysRevD.109.054011",
    journal = "Phys. Rev. D",
    volume = "109",
    number = "5",
    pages = "054011",
    year = "2024"
}

@article{Zhao:2023nrz,
    author = "Zhao, Jiaxing and others",
    title = "{Hadronization of heavy quarks}",
    eprint = "2311.10621",
    archivePrefix = "arXiv",
    primaryClass = "hep-ph",
    doi = "10.1103/PhysRevC.109.054912",
    journal = "Phys. Rev. C",
    volume = "109",
    number = "5",
    pages = "054912",
    year = "2024"
}

@article{Wigner:1932eb,
    author = "Wigner, Eugene P.",
    title = "{On the quantum correction for thermodynamic equilibrium}",
    doi = "10.1103/PhysRev.40.749",
    journal = "Phys. Rev.",
    volume = "40",
    pages = "749--760",
    year = "1932"
}

@article{Hillery:1983ms,
    author = "Hillery, M. and O'Connell, R. F. and Scully, M. O. and Wigner, Eugene P.",
    title = "{Distribution functions in physics: Fundamentals}",
    doi = "10.1016/0370-1573(84)90160-1",
    journal = "Phys. Rept.",
    volume = "106",
    pages = "121--167",
    year = "1984"
}

@article{Case:2008ped,
       author = "Case, William B.",
        title = "Wigner functions and Weyl transforms for pedestrians",
        doi = "10.1119/1.2957889",
        journal = "American Journal of Physics",
        volume = "76",
        number = "10",
        pages = "937-946",
        year = "2008"
}

\end{document}